\documentclass[11pt,preprintnumbers,amsmath,amssymb,nofootinbib]{revtex4}
\usepackage{exscale}
\usepackage{amsmath}
\usepackage{booktabs}
\usepackage{graphicx}
\usepackage{exscale}
\usepackage{relsize}
\usepackage{graphicx}
\usepackage{dcolumn}
\usepackage{bm}
\usepackage{multirow}
\usepackage{CJK}
\usepackage{diagbox}
\usepackage{subfigure}
\usepackage{slashed}
\usepackage{makecell}
\usepackage{adjustbox}

\def\Re{\text{Re}}
\def\Im{\text{Im}}

\begin{document}

\title{A comprehensive study on the semileptonic decay of heavy flavor mesons }

\author{Lu Zhang$^{1}$, Xian-Wei Kang$^{1,2}$, Xin-Heng Guo$^{1}$, Ling-Yun Dai$^{3}$, Tao Luo$^{4}$, Chao Wang$^5$}
\affiliation{{$^{1}$ Key Laboratory of Beam Technology of Ministry of Education, College of Nuclear Science and Technology, Beijing Normal University, Beijing 100875, People's Republic of China\\
$^{2}$ Beijing Radiation Center, Beijing 100875, China\\
$^{3}$ School of Physics and Electronics, Hunan University, Changsha 410082, China\\
$^{4}$ Key Laboratory of Nuclear Physics and Ion-beam Application (MOE) and Institute of Modern Physics,
Fudan University, Shanghai 200443, China\\
$^5$   School of Ecology and Environment, Northwestern Polytechnical University, Xi'an 710072, China
}}

\begin{abstract}
The semileptonic decay of heavy flavor mesons offers a clean environment for extraction of the Cabibbo-Kobayashi-Maskawa (CKM) matrix elements, which describes the CP-violating and flavor changing process in the Standard Model.
The involved form factors where the dynamical information is encoded play an essential role in achieving any conclusive statement. That is, the knowledge of the form factors should be under good control, requiring one to examine more observables in addition to the branching fraction.
In this paper, we provide the mean value and the $q^2$-dependent shape for further observables [differential decay distribution ($d\Gamma/dq^2$), forward-backward asymmetry ($\mathcal{A}_{FB}^l$), longitudinal ($P_L^l$) and transverse ($P_T^l$) polarization of a charged lepton, longitudinal polarization of a vector meson in the final state ($F_L^l(V)$), leptonic convexity parameter ($C_F^l$), and trigonometric moments ($W_i^l$) in the decay of $D_{(s)}$ and $B_{(s)}$ to $P/V l^+ \nu_l$ ($l=e$,\,$\mu$ or $\tau$)], based on the predictions of the relevant form factors from the covariant light-front quark model.
$P$ and $V$ denote the pseudoscalar and vector meson, respectively.
As a comparison, we find a good agreement with the results from the covariant confining quark model and the relativistic quark model in the literature.
As it has been observed that the $P_L^l$ and $F_L^l(V)$ are crucial quantities to discriminate various New Physics models, the reexamination of these observables from a different method is also essential and necessary.
\end{abstract}

\maketitle

\section{INTRODUCTION}
In charged-current transition, the semileptonic decay of heavy flavor mesons offers a clean environment (compared to the pure hadronic decay) to determine the Cabibbo-Kobayashi-Maskawa (CKM) matrix elements, which describes the quark flavor mixing \cite{Cabibbo:1963yz,Kobayashi:1973fv} in the Standard Model (SM). The precise determination of the matrix elements is one of the main tasks for both theorists and experimentalists. Any deviation from the unitarity relation encoded in the CKM matrix will certainly be an inspiring signal of New Physics (NP). As a result, the semileptonic decays are comprehensively studied to test the SM, or instead, to constrain NP parameters. In fact, there exist tensions for both $|V_{ub}|$ and $|V_{cb}|$ from their respective determinations by inclusive and exclusive modes \cite{Kang:2013jaa,Tanabashi:2018oca,Amhis:2019ckw}. For a most recent review, see Ref. \cite{Gambino:2020jvv}. Another aspect of hunting for NP via the semileptonic channel concerns the lepton-flavor universality violation; see the most recent review in \cite{Bifani:2018zmi}. To gain a deeper understanding of these issues, re-examination of the semileptonic channel is not obsolete but rather is worth doing again.

The hadronic and leptonic currents are involved in the semileptonic decay, and the former is parameterized by Lorentz-invariant form factors containing the dynamical information. The theoretical precision of this channel relies on our good control of the knowledge of form factors. In this paper, we exploit the results from the covariant light-front quark model (CLFQM) (Ref.~\cite{Cheng:2003sm} and updated in Ref.~\cite{Verma:2011yw}). The light-front form was initiated by Dirac in a formulation of Hamiltonian dynamics \cite{Dirac:1949cp} and later developed and applied to the hadron transition in Refs.~\cite{Jaus:1989au, Jaus:1991cy,Choi:1999nu}.
However, owing to the absence of a zero-mode effect, the conventional light-front quark model is noncovariant. To compensate for this deficiency, the CLFQM was proposed \cite{Jaus:1999zv} (some recent applications can be found, e.g., Refs.~\cite{Chang:2019mmh,Chang:2020wvs,Zhu:2018jet,Ke:2019smy,Shi:2016gqt}), where all four-momentum of each vertex is conserved, and the constituent quarks are off-shell. In Ref.~\cite{Cheng:2003sm}, the authors fixed the free parameters $\beta$ in the Gaussian-type wave function by the input value of the decay constant from other theory predictions, lattice and/or experiment results and then predicted the decay form factors. In Refs.~\cite{Cheng:2017pcq, Kang:2018jzg}, the authors calculated the corresponding branching ratios of some relevant decay channels, forming a direct bridge between phenomenological calculation and experiment. The result shows that the form factors given in Ref.~\cite{Cheng:2003sm,Verma:2011yw} nicely reproduce the experimental values.

We stress that the branching fraction is not the whole landscape and is only a number. More observables are needed
to distinguish various model predictions. The correlations among a series of observables are crucial to test a model calculation. We elucidate this point in two examples.
One example concerns the $K\pi$ form factor in $\tau$ decay. Currently, the most sophisticated description of the form factors corresponds to the dispersive representation with imposed chiral symmetry and short-distance QCD behavior, e.g., in Refs.~\cite{Bernard:2013jxa,Jamin:2006tj}. However, other options are available, e.g., in Ref.~\cite{Kimura:2014wsa} and Belle paper \cite{Epifanov:2007rf}, where in the former reference the form factors are calculated up to one-loop using the Lagrangian, including vector mesons. The shape in Ref.~\cite{Kimura:2014wsa} differs dramatically from that in Ref.~\cite{Bernard:2013jxa}, which may be attributed to the perturbative versus nonperturbative treatment. In the experimental study of Ref. \cite{Epifanov:2007rf}, the superposition of Breit-Wigner functions is used in a pragmatic way, which violates the Watson final-state-interaction theorem and also the unitarity.
However, all these variants of the form factor describe equally well the event distribution data, clearly indicating that one may need deeper theoretical founds or test more observable predictions. The other example is related to antinucleon-nucleon interaction. We will demonstrate this example by the J\"ulich potential \cite{Mull:1994gz}, where one sees that the cross section is well reproduced by various models there (models A(Box), C, D), but the predictions on the polarizations and spin observables differ dramatically. This situation also holds for the interaction constructed in chiral effective field theory (ChEFT) \cite{Kang:2013uia,Dai:2017ont}. Such a potential was applied to various hadron processes, e.g., in $e^+e^-\to p \bar p$ \cite{Haidenbauer:2014kja} and $J/\psi\to \gamma p\bar p$ \cite{Kang:2015yka,Dai:2018tlc}. There, and especially in Ref.~\cite{Kang:2015yka}, both the model A(OBE) and that for ChEFT successfully reproduce the phase shift (on-shell behavior) in the partial wave of $^1S_0$; however, they predict very different decay amplitudes in the distorted-wave Born approximation that involves the off-shell $T$-matrix elements. In a word, probing other observables apart from the branching fraction (integrated rate) is crucial to examine the theoretical models, which drives us to access the differential decay rate and the polarization.

In this paper, we will present such observables mentioned above in the semileptonic decays of heavy flavor mesons ($D,\,D_s,\,B,\,B_s$), namely,
$D_{(s)}$, $B_{(s)}$ decaying to $P l^+ \nu_l$ and $V l^+ \nu_l$ ($l=e$, $\mu$, or $\tau$), with $P$ and $V$ denoting the pseudoscalar and vector meson, respectively. For the ease of the reader, and to define our notation consistently, we closely follow Ref.~\cite{Ivanov:2019nqd} and write out the relevant equations\footnote{We perceive that the helicity formalism may be elegant. One example is that the decay rate for the semileptonic decay of a baryon, where the helicity formalism is analyzed in Ref.~\cite{Gutsche:2015mxa}, and the direct calculation is given in Ref.~\cite{Garcia:1992qe}. The former is mathematically more straightforward than the latter.}. In such a framework, one can evaluate the hadronic
tensor in the mother-particle (e.g., $D$) rest frame but the leptonic tensor in the rest frame of a $W$ boson, avoiding the somewhat complicated
boost. Then, the observables are expressed by the combination of form factors via helicity amplitudes, including the differential decay rate, forward-backward asymmetry (denoted by $A_{FB}^l$), longitudinal ($P_L^l$) and transverse ($P_T^l$) polarization of the charged lepton ($P_L^l$),  longitudinal polarization of vector in the final state ($F_L^l(V)$), and trigonometric moments ($W_i^l$).

Additionally, we notice that the observables of $A_{FB}^l$, $P_L^l$ have already been calculated in the covariant confining quark model (CCQM) \cite{Soni:2018adu,Ivanov:2015tru} and the relativistic quark model (RQM) \cite{Faustov:2019mqr}. From a theoretical viewpoint, we need to complete the corresponding calculation in light of the CLFQM as an indispensable complement and comparison. In particular, some of the observables have been measured in the experiment, which will be used to rectify our theoretical predictions.
These constituents are another
motivation for the current work, continuing the series of our previously published papers \cite{Cheng:2017pcq,Kang:2018jzg,Faustov:2019mqr,Cheng:2017qpv}.

In this work, we will consider the prediction of various observables, as mentioned above, based on the form factors calculated in the CLFQM. A full comparison
will be made with the ones from CCQM and RQM, wherever available. Here we add some comments on the similarity and difference between these three models. They are all a kind of quark model and thus share some common features, e.g., containing the wave function (bound state property) of initial and final state mesons, the quark masses as parameters. Some differences are in the treatments of wave functions.  The CCQM \cite{Ivanov:2015tru} is based on the effective Lagrangian describing the coupling of hadron to its constituent quarks. The coupling constant is fixed by saturating the Weinberg compositeness to guarantee the meson as a pure bound state of a quark and antiquark. For the vertex function, a simple Gaussian function is adopted with cutoff $\Lambda$ characterizing
the finite size of meson. The CLFQM \cite{Cheng:2003sm} is based on Dirac's light-front Hamiltonian dynamics and the relevant variables are expressed in the light-front system. The light-front vertex function derived from the Melosh transformation encodes the bound state nature. In the RQM~\cite{Faustov:2019mqr}, the wave function can be obtained by solving the quasipotential equation numerically, not necessary to assume a Gaussian  function. In addition, the form factors in the CCQM and RQM can be calculated in the whole $q^2$ region while CLFQM usually predicts the form factor in the spacelike region and then one analytically continue it to the (physical) timelike region.

This paper is organized as follows. In Sec.~\ref{Sec:ff}, we introduce the form factors and helicity amplitudes of the transitions $D(B)_{(s)}\rightarrow Pl^+\nu_l$ and $D(B)_{(s)}\rightarrow Vl^+\nu_l$. In Sec.~\ref{Sec:dGamma}, we provide the differential decay distribution, expressed by the helicity structure functions, while in Sec.~\ref{Sec:observable} the definitions of various observables, as mentioned above, are presented. In Sec.~\ref{Sec:results}, we give the numerical results of these physical observables in their mean value and their $q^2$ spectrum. We also compare the results with those from the constituent confining quark model \cite{Ivanov:2019nqd} and the relativistic quark model \cite{Faustov:2019mqr}, and excellent agreement is obtained. We conclude in Sec.~\ref{Sec:summary}.

\section{FORM FACTORS AND HELICITY AMPLITUDES}
\label{Sec:ff}

In this paper, we take the decay process of $D$ as an example to demonstrate our derivation for the relevant equations. For the $D_{s}$, $B$, and $B_{s}$ meson, just replacing some relevant physical parameters, the calculation is completely similar.

In the SM, the invariant matrix element for semileptonic decay of $D$ meson to a pseudoscalar ($P$) or a vector ($V$) meson can be written as
\begin{equation}
\begin{aligned}
\mathcal{M}\left(D\rightarrow P(V)l^+ \nu_l\right)&=\frac{G_F}{\sqrt{2}}V_{cq}\left\langle P(V)\mid \bar{q}O^\mu c\mid D_{(s)}\right\rangle l^+O_\mu \nu_l\\
&=\frac{G_F}{\sqrt{2}}V_{cq}H^{\mu}L_{\mu},
\end{aligned}
\end{equation}
where $O^\mu=\gamma^\mu(1-\gamma_5)$, $q=d,s$, and $G_F=1.166\times10^{-5}\,\text{GeV}^{-2}$ is the fermion coupling constant.

The hadronic part can be parameterized by the invariant form factors, which are a function of the momentum transfer squared $(q^2)$, between the initial and final meson. For the form factors, we stick to the commonly used Bauer-Stech-Wirbel (BSW) form \cite{Wirbel:1985ji}.

For the transition of $D\rightarrow Pl^+ \nu_l$, one has
\begin{equation}
\left\langle P(p_2)|V_\mu|D(p_1)\right\rangle=T_\mu,
\end{equation}
and the form factors are given by
\begin{equation}
T_\mu=\left(P_\mu-\frac{m_1^2-m_2^2}{q^2}q_\mu\right)F_1(q^2)+\frac{m_1^2-m_2^2}{q^2}q_\mu F_0(q^2).
\label{eqtp}
\end{equation}
For the transition of $D\rightarrow Vl^+ \nu_l$, one has
\begin{equation}
\left\langle V(p_2)|V_\mu-A_\mu|D(p_1)\right\rangle=\epsilon_2^{\dagger\nu} T_{\mu\nu},
\end{equation}
where
\begin{equation}
\begin{split}
T_{\mu\nu}&=-\left(m_1+m_2\right)\left[g_{\mu\nu}-\frac{P_\nu}{q^2}q_\mu\right]A_1(q^2) +\frac{P_\nu}{m_1+m_2}\left[P_\mu-\frac{m_1^2-m_2^2}{q^2}q_\mu\right]A_2(q^2)\\
&-2m_2\frac{P_\nu}{q^2}q_\mu A_0(q^2) +\frac{i}{m_1+m_2}\varepsilon_{\mu\nu\alpha\beta} P^\alpha q^\beta V(q^2).
\end{split}
\label{eqtv}
\end{equation}
In the above equations, $p_1$ ($m_1$) and $p_2$ ($m_2$) are the four-momentum (mass) of the initial and final meson, respectively. The momenta $P_\mu$ and $q_\mu$ are defined as $P_\mu={(p_1+p_2)}_\mu$ and $q_\mu={(p_1-p_2)}_\mu$, respectively. $\epsilon_2$ is the polarization vector of the final state meson $V$ and satisfies $\epsilon_2^\dagger \cdot p_2=0$.

In the covariant contraction of hadronic and leptonic tensors, $H^{\mu\nu}L_{\mu\nu}$,
the hadronic tensor can be written as
\begin{equation}
\begin{split}
H^{\mu\nu}&=\sum\left\langle X|V^\mu-A^\mu|D\right\rangle\left\langle X|V^\nu-A^\nu|D\right\rangle^\dagger\\
&=\begin{cases}T_\mu T_\nu^\dagger, \quad\quad\quad\quad\quad\quad\quad\quad\quad\quad D\rightarrow P,\\
T_{\mu\alpha}T_{\nu\beta}^\dagger\left(-g^{\alpha\beta}+\frac{p_2^\alpha p_2^\beta}{m_2^2}\right),\quad\quad D\rightarrow V.
\end{cases}
\end{split}
\end{equation}
We can define the explicit representations of polarization four-vectors $\epsilon^\mu(\lambda_W)$, where $\lambda_W$ denotes the polarization index of $W_\text{off-shell}$, and $\epsilon^\mu(\lambda_W)q_\mu=0$ for $\lambda_W=\pm,0$. These four-vectors $\epsilon^\mu(\lambda_W)$ have the orthonormality property
\begin{equation}
\epsilon_\mu^\dagger(\lambda_W)\epsilon^\mu({\lambda^{\prime}_W})=g_{\lambda_W{\lambda^{\prime}_W}},\quad (\lambda_W,{\lambda^{\prime}_W}=t,\pm,0)
\end{equation}
and satisfy the completeness relation
\begin{equation}
\epsilon_\mu(\lambda_W)\epsilon_\nu^\dagger({\lambda^{\prime}_W})g_{\lambda_W{\lambda^{\prime}_W}}=g_{\mu\nu}.
\end{equation}
We can rewrite the contraction of leptonic and hadronic tensors by using the orthonormality and completeness relations as
\begin{equation}
\begin{split}
L^{\mu\nu}H_{\mu\nu}&=L_{\mu^\prime \nu^\prime}g^{\mu^\prime \mu}g^{\nu^\prime \nu}H_{\mu\nu}\\
&=L_{\mu^\prime \nu^\prime}\epsilon^{\mu^\prime}(\lambda_W)\epsilon^{\dagger\mu}(\lambda_W^{\prime\prime})
g_{\lambda_W\lambda_W^{\prime\prime}}\epsilon^{\dagger \nu^\prime}(\lambda_W^\prime)\epsilon^\nu(\lambda_W^{\prime\prime\prime})g_{\lambda_W^\prime\lambda_W^{\prime\prime\prime}}H_{\mu\nu}\\
&=L\left(\lambda_W,\lambda_W^{\prime}\right)g_{\lambda_W\lambda_W^{\prime\prime}}g_{\lambda_W^\prime\lambda_W^{\prime\prime\prime}}H\left(\lambda_W^{\prime\prime}\lambda_W^{\prime\prime\prime}\right),
\end{split}
\end{equation}
where $L(\lambda_W,\lambda_W^{\prime})$ and $H(\lambda_W,\lambda_W^{\prime})$ are the leptonic and hadronic tensors in the helicity-component space:
\begin{equation}
\label{eq:LHhelicity}
L\left(\lambda_W,\lambda_W^{\prime}\right)=\epsilon^\mu(\lambda_W)\epsilon^{\dagger\nu}(\lambda_W^{\prime})L_{\mu\nu},\quad H\left(\lambda_W,\lambda_W^{\prime}\right)=\epsilon^{\dagger\mu}(\lambda_W)\epsilon^\nu(\lambda_W^{\prime})H_{\mu\nu}.
\end{equation}
It is necessary to stress that the leptonic and hadronic tensor can be evaluated in two different Lorentz frames within this framework: the leptonic tensor is evaluated in the $W$ rest frame, and the hadronic tensor is evaluated in the $D$ rest frame.

\subsection{The helicity amplitudes of $D\rightarrow Pl^+\nu_l$}
For the transition $D\rightarrow Pl^+\nu_l$ in the helicity-component space, the hadronic tensor is given by
\begin{equation}
\begin{split}
H\left(\lambda_W,\lambda_W^{\prime}\right)&=\epsilon^{\dagger\mu}(\lambda_W)\epsilon^\nu(\lambda_W^{\prime})H_{\mu\nu}\\
&=[\epsilon^{\dagger\mu}(\lambda_W)T_\mu][\epsilon^{\dagger\nu}(\lambda_W^{\prime})T_\nu]^\dagger= H_{\lambda_W} H_{\lambda_W^{\prime}}^\dagger,
\end{split}
\label{eqhp}
\end{equation}
with $H_{\lambda_W}\equiv\epsilon^{\mu\dagger}(\lambda_W)T_\mu$.
In the rest frame of the initial meson $D$, one has the explicit representations \cite{Ivanov:2015tru} of the polarization four-vectors $\epsilon^\mu(\lambda_W)$:
\begin{equation}
\epsilon^\mu(t)=\frac{1}{\sqrt{q^2}}\left(q_0,0,0,|\vec{p}_2|\right),\quad \epsilon^\mu(\pm)=\frac{1}{\sqrt{2}}\left(0,\mp1,-i,0\right),\quad \epsilon^\mu(0)=\frac{1}{\sqrt{q^2}}\left(\left|\vec{p}_2\right|,0,0,q_0\right).
\label{eqwp}
\end{equation}
Among these representations, $q_0$ is the energy and $|\vec{p}_2|$ is the momentum of the $W_\text{off-shell}$ in the $D$ rest frame. Additionally, in this frame (see Fig.~\ref{angle}),
\begin{equation}
\begin{aligned}
&p_1^\mu=\left(m_1,0,0,0\right),\\
&p_2^\mu=\left(E_2,0,0,-|\vec{p}_2|\right),\\
&q^\mu=\left(q_0,0,0,|\vec{p}_2|\right),
\end{aligned}
\end{equation}
where $E_2$ is the energy of the final meson, and $q_0$ and $|\vec{p}_2|$ are given by
\begin{equation}
2m_1q_0=m_1^2-m_2^2+q^2,\quad
\left|\vec{p}_2\right|^2=E_2^2-m_2^2,\quad
2m_1E_2=m_1^2+m_2^2-q^2.\quad
\end{equation}

To calculate the relations between form factors and helicity amplitudes, we project out the relevant helicity amplitudes using the polarization four-vectors $\epsilon^\mu(\lambda_W)$. Combining Eq.~\eqref{eqtp} and Eq.~\eqref{eqwp}, for the $D\rightarrow P$ transition, we obtain
\begin{equation}
\begin{aligned}
&H_t=\frac{1}{\sqrt{q^2}}(m_1^2-m_2^2)F_0(q^2),\\
&H_\pm=0,\\
&H_0=\frac{2m_1|\vec{p}_2|}{\sqrt{q^2}}F_1(q^2).
\end{aligned}
\end{equation}
The invariant form factors are only a function of $q^2$. From this equation, we find that $H_t=H_0$ at the maximum recoil $(q^2=0)$, and $H_0=0$ at the zero recoil $(q^2=q^2_\text{max})$.

\subsection{The helicity amplitudes of $D\rightarrow Vl^+\nu_l$}
For the transition $D\rightarrow Vl^+\nu_l$, we also need the polarization four-vectors $\epsilon_2^\nu(\lambda_V)$ to obtain the expressions of the helicity amplitudes. The index $\lambda_V$ denotes the polarization index of vector meson $V$.
The hadronic tensor of the $D\rightarrow V$ transition can be rewritten as
\begin{equation}
\begin{split}
H\left(\lambda_W,\lambda_W^{\prime}\right)&=\epsilon^{\dagger\mu}(\lambda_W)\epsilon^\nu(\lambda_W^{\prime})H_{\mu\nu}\\
&=\epsilon^{\dagger\mu}(\lambda_W)\epsilon^\nu(\lambda_W^{\prime})\epsilon_2^{\dagger\alpha}(\lambda_V)T_{\mu\alpha}\left[\epsilon_2^{\dagger\beta}(\lambda_V^{\prime})T_{\nu\beta}\right]^\dagger\delta_{\lambda_V\lambda_V^{\prime}}\\
&=\epsilon^{\dagger\mu}(\lambda_W)\epsilon_2^{\dagger\alpha}(\lambda_V)T_{\mu\alpha}\left[\epsilon^{\dagger\nu}(\lambda_W^{\prime})\epsilon_2^{\dagger\beta}(\lambda_V^{\prime})T_{\nu\beta}\right]^\dagger\delta_{\lambda_V\lambda_V^{\prime}}\\
&=H_{\lambda_W\lambda_V}H_{\lambda_W^{\prime}\lambda_V}^\dagger,
\end{split}
\label{eqhv}
\end{equation}
with
\begin{equation}
H_{\lambda_W\lambda_V}\equiv\epsilon^{\dagger\mu}(\lambda_W)\epsilon_2^{\dagger\alpha}(\lambda_V)T_{\mu\alpha}.
\label{eqh}
\end{equation}
According to the angular momentum conservation, $\lambda_V=\lambda_W, \lambda_V^{\prime}=\lambda_W^{\prime}$ for $\lambda_W,\lambda_W^{\prime}=\pm,0$; and $\lambda_V,\lambda_V^{\prime}=0$ for $\lambda_W,\lambda_W^{\prime}=t$.

In the $D$ rest frame, the explicit representations of the polarization four-vectors for the final vector meson $\epsilon_2^\nu(\lambda_V)$ can be written as
\begin{equation}
\epsilon_2^\nu(\pm)=\frac{1}{\sqrt{2}}\left(0,\pm1,-i,0\right),\quad \epsilon_2^\nu(0)=\frac{1}{m_2}\left(|\vec{p}_2|,0,0,-E_2\right).
\end{equation}
They satisfy the orthonormality condition
\begin{equation}
\epsilon_{2\nu}^\dagger(\lambda_V)\epsilon_2^\nu({\lambda^{\prime}_V})=-\delta_{\lambda_V{\lambda^{\prime}_V}},
\end{equation}
and the completeness relation
\begin{equation}
\sum\epsilon_{2\mu}(\lambda_V)\epsilon_{2\nu}^\dagger({\lambda^{\prime}_V})\delta_{\lambda_V{\lambda^{\prime}_V}}=-g_{\mu\nu}+\frac{p_{2\mu}p_{2\nu}}{m_2^2}.
\end {equation}

In terms of the Eq.~\eqref{eqh}, we use a similar method to obtain the nonvanishing helicity amplitudes of the transition $D\rightarrow Vl^+\nu_l$:
\begin{equation}
\begin{aligned}
H_t&\equiv\epsilon^{\dagger\mu}(t)\epsilon_2^{\dagger\nu}(0)T_{\mu\nu}=-\frac{2m_1|\vec{p}_2|}{\sqrt{q^2}}A_0(q^2),\\
H_\pm&\equiv\epsilon^{\dagger\mu}(\pm)\epsilon_2^{\dagger\nu}(\pm)T_{\mu\nu}=-(m_1+m_2)A_1(q^2)\pm\frac{2m_1|\vec{p}_2|}{m_1+m_2}V(q^2),\\
H_0&\equiv\epsilon^{\dagger\mu}(0)\epsilon_2^{\dagger\nu}(0)T_{\mu\nu}=-\frac{m_1+m_2}{2m_2\sqrt{q^2}}\left(m_1^2-m_2^2-q^2\right)A_1(q^2)+\frac{1}{m_1+m_2}\frac{2m_1^2|\vec{p}_2|^2}{m_2\sqrt{q^2}}A_2(q^2).
\end{aligned}
\end{equation}
At the maximum recoil ($q^2=0$), we have $H_t(0)=H_0(0)$, and $H_t$ and $H_0$ dominate, while at the zero recoil ($q^2=q^2_\text{max}$), $H_t=0$ and $H_{\pm}=H_0$.

\subsection{The hadronic tensor including the cascade decay $V\rightarrow P_1P_2$}
\begin{figure}[ht]
\begin{minipage}{1.0\linewidth}
\centerline{\includegraphics[width=0.3\textwidth]{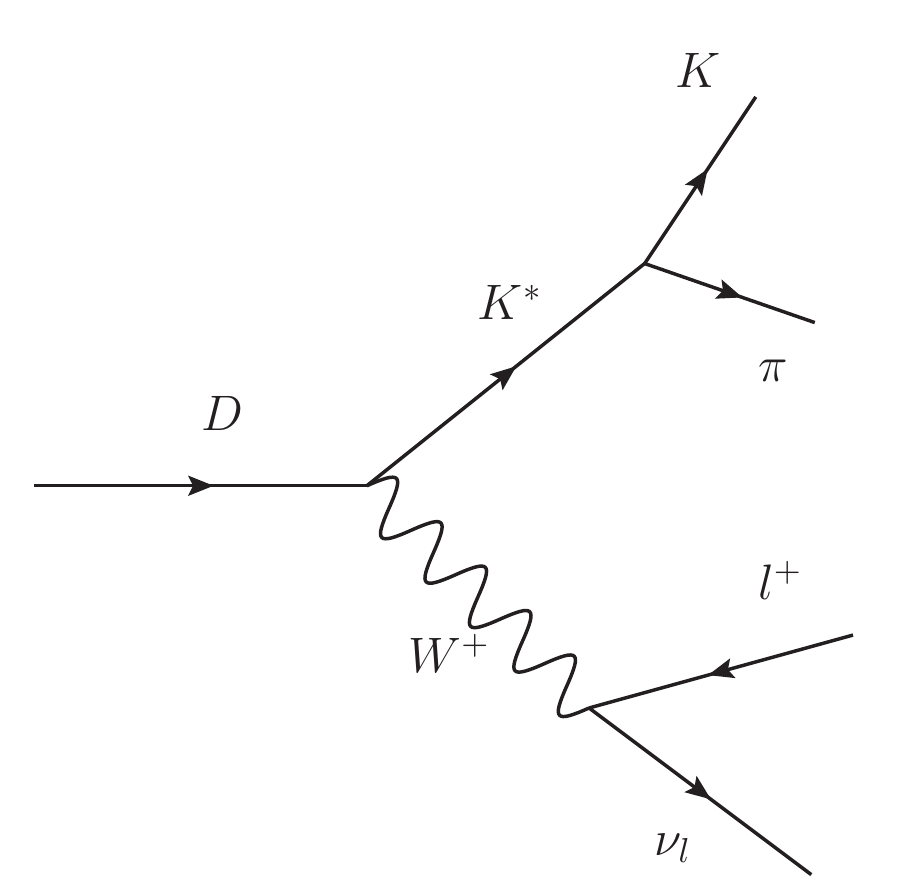}}
\end{minipage}
\caption{Illustration of the semileptonic decay process $D\rightarrow K^*(\rightarrow K\pi)l\nu_l$.}
\label{figtran}
\end{figure}

Taking into account the finite width effect of the vector meson, we need to consider the cascade decay $V\rightarrow P_1P_2$ with $P$ denoting a pseudoscalar. For a more intuitive description, we give an example of cascade decay in Fig.~\ref{figtran}.

\begin{figure}[ht]
\begin{minipage}{1.0\linewidth}
\centerline{\includegraphics[width=0.6\textwidth]{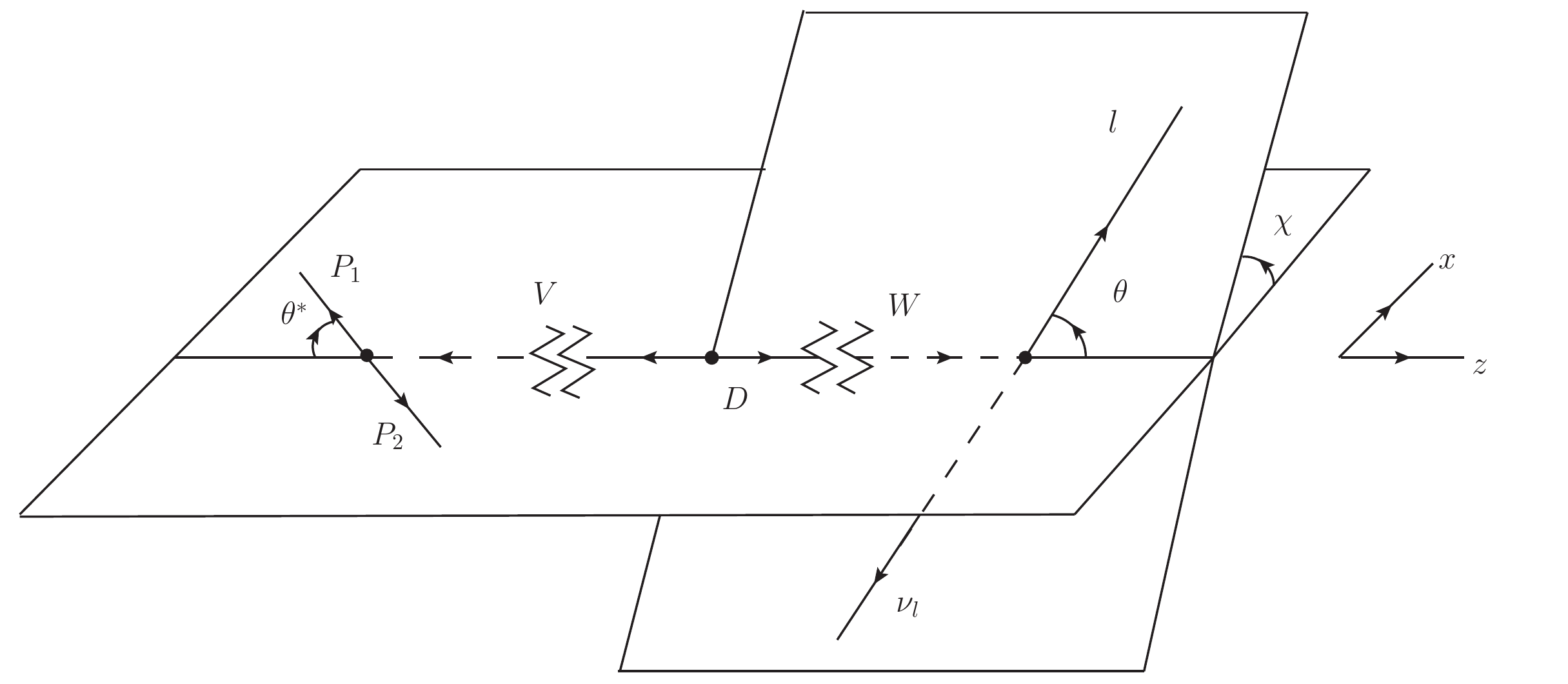}}
\end{minipage}
\caption{The definitions of $\theta$, $\theta^*$ and the azimuthal angle $\chi$.}
\label{angle}
\end{figure}

The kinematics for four-body decay is elaborated in Refs.~\cite{Kang:2013jaa,Cheng:2017qpv,Stoffer:2013sfa} and earlier papers \cite{Cabibbo:1965zzb,Pais:1968zza}. As shown in Fig.~\ref{angle},
we define $\theta$ as the polar angle of the lepton in the $l^+\nu_l$ center-of-mass frame with respect to the $W_\text{off-shell}$ line of flight in the $D$ rest frame.
Additionally, $\theta^*$ is defined as the polar angle of $P_1$ in the $V$ rest frame with respect to the $V$ meson line of flight in the $D$ rest frame.
$\chi$ is the dihedral angle between the $l^+\nu_l$ and the $P_1P_2$ planes. The $z$-axis is along the momentum direction of $W_\text{off-shell}$ in the $D$ rest frame, and the momenta of $P_1$ and $P_2$ lie in the $xz$ plane.

In the rest frame of final meson $V$, the corresponding momenta are
\begin{equation}
\begin{aligned}
p_2^\mu&=\left(m_2,0,0,0\right),\\
p_3^\mu&=\left(E_3,|\vec{p}_3|\sin\theta^*,0,-|\vec{p}_3|\cos\theta^*\right),\\
p_4^\mu&=\left(E_4,-|\vec{p}_3|\sin\theta^*,0,|\vec{p}_3|\cos\theta^*\right),
\label{eq:Vrest}
\end{aligned}
\end{equation}
and the polarization vectors of $V$ meson are
\begin{equation}
\epsilon_2^\mu(+)=\frac{1}{\sqrt{2}}\left(0,+1,-i,0\right),\; \epsilon_2^\mu(-)=\frac{1}{\sqrt{2}}\left(0,-1,-i,0\right),\; \epsilon_2^\mu(0)=\left(0,0,0,-1\right).
\label{eqpol2}
\end{equation}

In the decay process of $D\to V (\to P_1P_2) l\nu_l$, the hadronic tensor is given by
\begin{equation}
\begin{aligned}
H\left(\lambda_W,\lambda_W^{\prime}\right)&=\epsilon^{\dagger\mu}(\lambda_W)\epsilon^\nu(\lambda_W^{\prime})H_{\mu\nu}\\
&=\epsilon^{\dagger\mu}(\lambda_W)\epsilon^\nu(\lambda_W^{\prime})\epsilon_2^{\dagger\alpha}(\lambda_V)\epsilon_2^{\beta}(\lambda_V^{\prime})
T_{\mu\alpha}(T_{\nu\beta})^\dagger g^2_{VPP}p_{3\alpha'}p_{3\beta'}\epsilon_2^{\alpha'}(\lambda_V)\epsilon_2^{\dagger\beta'}(\lambda_V^{\prime})\\
&=g^2_{VPP}p_{3\alpha'}p_{3\beta'}\epsilon_2^{\alpha'}(\lambda_V)\epsilon_2^{\dagger\beta'}(\lambda_V^{\prime})\times H_{\lambda_W,\lambda_V}H^{\dagger}_{\lambda_W^{\prime},\lambda_V^{\prime}},
\end{aligned}
\label{eq:TensorCascade}
\end{equation}
where $g_{VPP}\cdot \epsilon_2^{\rho}(\lambda_V) \cdot p_{3\rho}$ describes the amplitude $\mathcal{A}(V\to P_1P_2)$, and the effective coupling constant $g_{VPP}$
is related to the branching ratio $Br(V\to P_1P_2)$ by
\begin{equation}
g^2_{VPP}=\frac{24\pi m_{2}^2 \Gamma_{V} Br}{|\vec{p}_3|^3},
\end{equation}
with $\Gamma_V$ being the finite width of vector meson.
Notice that $S^{\alpha\alpha'}(p_2)=\displaystyle\sum_{\lambda_V=\pm1,0}\epsilon_2^{\alpha}(\lambda_V)\epsilon_2^{\dagger\alpha'}(\lambda_V)$ is just the numerator of the propagator of intermediate meson $V$. Additionally, Eq. \eqref{eq:TensorCascade} agrees with Eq.~(37) in Ref.~\cite{Ivanov:2019nqd},
except for some extra factors that will be compensated in the final result.

\section{THE TWOFOLD AND FOURFOLD DIFFERENTIAL DECAY DISTRIBUTIONS}
\label{Sec:dGamma}
In this section, we consider the differential decay distribution.
For the $D\rightarrow P(V)l^+\nu_l$ transition, we have
\begin{equation}
\begin{split}
\frac{d^2\Gamma(D\rightarrow P(V)l^+\nu_l)}{dq^2d\cos\theta}&=\frac{|\vec{p}_2|}{(2\pi)^3 32m_1^2}\left(1-\frac{m_l^2}{q^2}\right)\sum\limits_{\text{spins}}\left|\mathcal{M}\right|^2\\
&=\frac{G_F^2|V_{cq}|^2}{(2\pi)^3}\frac{|\vec{p}_2|}{64m_1^2}\left(1-\frac{m_l^2}{q^2}\right)
L\left(\lambda_W,\lambda_W^{\prime}\right)g_{\lambda_W\lambda_W^{\prime\prime}}g_{\lambda_W^{\prime}\lambda_W^{\prime\prime\prime}}H\left(\lambda_W^{\prime\prime},\lambda_W^{\prime\prime\prime}\right),
\end{split}
\end{equation}
where $m_l$ is the lepton mass, and the factor in front of $|\mathcal{M}|^2$ corresponds to the three-body phase factor. In terms of the helicity amplitudes, Eqs.~\eqref{eqhp} and~\eqref{eqhv}, the hadronic tensor is
\begin{equation}
H\left(\lambda_W,\lambda_W^{\prime}\right)=\begin{cases}H_{\lambda_W} H_{\lambda_W^{\prime}}^\dagger, \quad\quad\quad\quad D\rightarrow P,\\
H_{\lambda_W\lambda_V}H_{\lambda_W^{\prime}\lambda_V}^\dagger,\quad\quad D\rightarrow V.
\label{eqhd}
\end{cases}
\end{equation}

We then evaluate the leptonic tensor $L(\lambda_W,\lambda_W^{\prime})=\epsilon^\mu(\lambda_W)\epsilon^{\dagger\nu}(\lambda_W^{\prime})L_{\mu\nu}$. The leptonic tensor can be given by
\begin{equation}
\begin{split}
L_{\mu\nu}&=\begin{cases}\text{tr}\left[\left(\slashed{k}_1+m_l\right)O_\mu\slashed{k}_2O_\nu\right],\quad\quad W^-\rightarrow l^-\bar{\nu}_l,\\
\text{tr}\left[\left(\slashed{k}_1-m_l\right)O_\nu\slashed{k}_2O_\mu\right],\quad\quad W^+\rightarrow l^+\nu_l,
\end{cases}\\
&=8\left(k_{1\mu}k_{2\nu}+k_{1\nu}k_{2\mu}-k_1\cdot k_2g_{\mu\nu}\pm i\varepsilon_{\mu\nu\alpha\beta}k_1^\alpha k_2^\beta\right),
\end{split}
\label{eql}
\end{equation}
where $k_1,k_2$ denote the momentum of a charged lepton and neutrino, respectively. The evaluation of the leptonic tensor is performed in the $W_\text{off-shell}$ rest frame, and the explicit expressions of the momentum can be written as
\begin{equation}
\begin{split}
&q^\mu=\left(\sqrt{q^2},0,0,0\right),\\
&k_1^\mu=\left(E_1,|\vec{k}_1|\sin\theta \cos\chi,|\vec{k}_1|\sin\theta \sin\chi,|\vec{k}_1|\cos\theta\right),\\
&k_2^\mu=\left(|\vec{k}_1|,-|\vec{k}_1|\sin\theta \cos\chi,-|\vec{k}_1|\sin\theta \sin\chi,-|\vec{k}_1|\cos\theta\right),
\end{split}
\end{equation}
where $E_1=\left(q^2+m_l^2\right)/2\sqrt{q^2}$ is the energy and $|\vec{k_1}|=\left(q^2-m_l^2\right)/2\sqrt{q^2}$ is the three-momentum of the lepton. The polarization vectors of $W$ in this rest frame are given by
\begin{equation}
\epsilon^\mu(0)=\left(0,0,0,1\right),\quad\epsilon^\mu(\pm)=\frac{1}{\sqrt{2}}\left(0,\mp1,-i,0\right),\quad\epsilon^\mu(t)=\left(1,0,0,0\right).
\label{eqvw}
\end{equation}

Combining the momentum and polarization vectors, Eqs.~\eqref{eql}-\eqref{eqvw}, we obtain the expression for $L(\lambda_W,\lambda_W^{\prime})$, where the matrix columns and rows are ordered in the sequence $(t,+,0,-)$ (In Ref.~\cite{Ivanov:2019nqd}, the element in the first row and second column of the matrix after $\delta$ contains a typo, $\chi$ should be replaced by $\chi$):
\begin{equation}
\begin{aligned}
(2q^2v)^{-1}L(\lambda_W,\lambda_W^{\prime})(\theta,\chi)&=\left(\begin{array}{cccc}
0&0&0&0\\
0&(1\mp\cos\theta)^2&\mp\frac{2}{\sqrt{2}}(1\mp\cos\theta)\sin\theta e^{i\chi}&\sin^2\theta e^{2i\chi}\\
0&\mp\frac{2}{\sqrt{2}}(1\mp\cos\theta)\sin\theta e^{-i\chi}&2\sin^2\theta&\mp\frac{2}{\sqrt{2}}(1\pm\cos\theta)\sin\theta e^{i\chi}\\
0&\sin^2\theta e^{-2i\chi}&\mp\frac{2}{\sqrt{2}}(1\pm\cos\theta)\sin\theta e^{-i\chi}&(1\pm\cos\theta)^2
\end{array}\right)\\
&+\delta_l\left(\begin{array}{cccc}
4&-\frac{4}{\sqrt{2}}\sin\theta e^{-i\chi}&4\cos\theta&\frac{4}{\sqrt{2}}\sin\theta e^{i\chi}\\
-\frac{4}{\sqrt{2}}\sin\theta e^{i\chi}&2\sin^2\theta&-\frac{2}{\sqrt{2}}\sin2\theta e^{i\chi}&-2\sin^2\theta e^{2i\chi}\\
4\cos\theta&-\frac{2}{\sqrt{2}}\sin2\theta e^{-i\chi}&4\cos^2\theta&\frac{2}{\sqrt{2}}\sin2\theta e^{i\chi}\\
\frac{4}{\sqrt{2}}\sin\theta e^{-i\chi}&-2\sin^2\theta e^{-2i\chi}&\frac{2}{\sqrt{2}}\sin2\theta e^{-i\chi}&2\sin^2\theta
\end{array}\right).
\end{aligned}
\end{equation}
where the upper/lower sign refers to the $(l^-\bar{\nu}_l)/(l^+\nu_l)$ configuration. To obtain the distribution on the polar angular $\theta$, we integrate over the azimuthal angel $\chi$, and then, the matrix becomes
\begin{equation}
(2q^2v)^{-1}L(\lambda_W,\lambda_W^{\prime})(\theta)=\left(\begin{array}{cccc}
0&0&0&0\\
0&(1\mp \cos\theta)^2&0&0\\
0&0&2\sin^2\theta&0\\
0&0&0&(1\pm\cos\theta)^2
\end{array}\right)
+\delta_l\left(\begin{array}{cccc}
4&0&4\cos\theta&0\\
0&2\sin^2\theta&0&0\\
4\cos\theta&0&4\cos^2\theta&0\\
0&0&0&2\sin^2\theta
\end{array}\right),
\label{eqlm}
\end{equation}
where $v=1-m_l^2/q^2$ is the velocity-type parameter and $\delta_l=m_l^2/2q^2$ is the helicity-flip factor.

We consider now the circumstance of $W^+\rightarrow l^+\nu_l$, and the $W^-\rightarrow l^-\bar{\nu}_l$ case will be discussed later. Combing the Eq.~\eqref{eqlm} and Eq.~\eqref{eqhd}, one has the contraction $H^{\mu\nu}L_{\mu\nu}(\theta)$ in helicity amplitude space as follows:
\begin{equation}
\begin{split}
H^{\mu\nu}L_{\mu\nu}(\theta)=&L\left(\lambda_W,\lambda_W^{\prime}\right)g_{\lambda_W\lambda_W^{\prime\prime}}g_{\lambda_W^{\prime}\lambda_W^{\prime\prime\prime}}H\left(\lambda_W^{\prime\prime},\lambda_W^{\prime\prime\prime}\right)\\
=&\left(2q^2v\right)\times\left\lbrace\left[\left(1+\cos^2\theta\right)+2\delta_l\sin^2\theta\right]\left(\left|H_+\right|^2+\left|H_-\right|^2\right)
+\left(2\sin^2\theta+4\delta_l\cos^2\theta\right)\left|H_0\right|^2\right.\\
&\left.+2\cos\theta\left(\left|H_+\right|^2-\left|H_-\right|^2\right)+4\delta_l\left|H_t\right|^2-8\delta_l\cos\theta\, \text{Re} \left(H_0H_t^\dagger\right)\right\rbrace\\
=&\left(2q^2v\right)\times\left[\left(1+\cos^2\theta\right)\mathcal{H}_U+2\sin^2\theta\mathcal{H}_L + 2\cos\theta\mathcal{H}_P\right.\\
&\left.+2\delta_l\left(\sin^2\theta\mathcal{H}_U+2\cos^2\theta\mathcal{H}_L+2\mathcal{H}_S-4\cos\theta\mathcal{H}_{SL}\right)\right].
\end{split}
\end{equation}
The above equation contains relevant bilinear combinations $\mathcal{H}_i$, called by the helicity structure functions~\cite{Ivanov:2015tru}, whose definitions are listed in Table $\mathrm{\ref{tabh}}$.

\begin{table}[!htbp]
\centering
\caption{Definitions of helicity structure functions}
\begin{tabular}{cc}
\hline
\hline
Parity-conserving&Parity-violating\\
\hline
$\mathcal{H}_U=|H_+|^2+|H_-|^2$&$\mathcal{H}_P=|H_+|^2-|H_-|^2$\\
$\mathcal{H}_L=|H_0|^2$&$\mathcal{H}_A=\frac{1}{2}\Re(H_+H_0^\dagger-H_-H_0^\dagger)$\\
$\mathcal{H}_T=\Re(H_+H_-^\dagger)$&$\mathcal{H}_{IA}=\frac{1}{2}\Im(H_+H_0^\dagger-H_-H_0^\dagger)$\\
$\mathcal{H}_{IT}=\Im(H_+H_-^\dagger)$&$\mathcal{H}_{SA}=\frac{1}{2}\Re(H_+H_t^\dagger-H_-H_t^\dagger)$\\
$\mathcal{H}_I=\frac{1}{2}\Re(H_+H_0^\dagger+H_-H_0^\dagger)$&$\mathcal{H}_{ISA}=\frac{1}{2}\Im(H_+H_t^\dagger-H_-H_t^\dagger)$\\
$\mathcal{H}_{II}=\frac{1}{2}\Im(H_+H_0^\dagger+H_-H_0^\dagger)$&\;\\
$\mathcal{H}_{S}=|H_t|^2$&\;\\
$\mathcal{H}_{ST}=\frac{1}{2}\Re(H_+H_t^\dagger+H_-H_t^\dagger)$&\;\\
$\mathcal{H}_{IST}=\frac{1}{2}\Im(H_+H_t^\dagger+H_-H_t^\dagger)$&\;\\
$\mathcal{H}_{SL}=\Re(H_0H_t^\dagger)$&\;\\
$\mathcal{H}_{ISL}=\Im(H_0H_t^\dagger)$&\;\\
\hline
$\mathcal{H}_{tot}=\mathcal{H}_U+\mathcal{H}_L+\delta_l\left(\mathcal{H}_U+\mathcal{H}_L+3\mathcal{H}_S\right)$&\;\\
\hline
\hline
\end{tabular}
\label{tabh}
\end{table}

Then, we obtain the twofold differential decay distribution on $q^2$ and $\cos\theta$:
\begin{equation}
\begin{split}
\frac{d\Gamma\left(D\rightarrow P(V)l^+\nu_l\right)}{dq^2d\cos\theta}=&\frac{G_F^2|V_{cq}|^2|\vec{p}_2|q^2v^2}{32(2\pi)^3m_1^2}\times\left[\left(1+\cos^2\theta\right)\mathcal{H}_U+2\sin^2\theta \mathcal{H}_L+2\cos\theta \mathcal{H}_P\right.\\
&\left.+2\delta_l\left(\sin^2\theta \mathcal{H}_U+2\cos^2\theta \mathcal{H}_L+2\mathcal{H}_S-4\cos\theta\mathcal{H}_{SL}\right)\right].
\end{split}
\label{eq:Gammathree}
\end{equation}
Further integrating over $\cos\theta$, the differential $q^2$ distribution will be
\begin{equation}
\begin{aligned}
\frac{d\Gamma\left(D\rightarrow P(V)l^+\nu_l\right)}{dq^2}=&\frac{G_F^2|V_{cq}|^2|\vec{p}_2|q^2v^2}{12(2\pi)^3m_1^2}
\times\mathcal{H}_{tot},
\end{aligned}
\label{equn}
\end{equation}
with $\mathcal{H}_{tot}=\mathcal{H}_U+\mathcal{H}_L+\delta_l\left(\mathcal{H}_U+\mathcal{H}_L+3\mathcal{H}_S\right)$.

For the cascade decays, we need to consider the fourfold distribution. Equation~\eqref{eq:TensorCascade} shows the hadronic tensor, including the cascade decay. Combining the momenta of $P_1$ Eq.~\eqref{eq:Vrest} and the polarization vectors of $V$ meson Eq.~\eqref{eqpol2}, the hadronic tensor in the helicity space can be written as
\begin{equation}
\begin{aligned}
\frac{1}{g^2_{VPP}|\vec p_3|^2}H\left(\lambda_W,\lambda_W^{\prime}\right)=&\left(\begin{array}{cccc}
\cos^2\theta^*|H_t|^2&\frac{1}{2\sqrt{2}}\sin2\theta^*H_tH_+^{\dagger}&\cos^2\theta^*H_tH_0^{\dagger}&-\frac{1}{2\sqrt{2}}\sin2\theta^*H_tH_-^{\dagger}\\
\frac{1}{2\sqrt{2}}\sin2\theta^*H_+H_t^{\dagger}&\frac{1}{2}\sin^2\theta^*|H_+|^2&\frac{1}{2\sqrt{2}}\sin2\theta^*H_+H_0^{\dagger}&-\frac{1}{2}\sin^2\theta^*H_+H_-^{\dagger}\\
\cos^2\theta^*H_0H_t^{\dagger}&\frac{1}{2\sqrt{2}}\sin2\theta^*H_0H_+^{\dagger}&\cos^2\theta^*|H_0|^2&-\frac{1}{2\sqrt{2}}\sin2\theta^*H_0H_-^{\dagger}\\
-\frac{1}{2\sqrt{2}}\sin2\theta^*H_-H_t^{\dagger}&-\frac{1}{2}\sin^2\theta^*H_-H_+^{\dagger}&-\frac{1}{2\sqrt{2}}\sin2\theta^*H_-H_0^{\dagger}&\frac{1}{2}\sin^2\theta^*|H_-|^2
\end{array}\right),
\end{aligned}
\end{equation}
with the matrix orders $\lambda_W,\lambda_W^{\prime}=t,+,0,-$.

The four-body phase space for $D\to V(\to P_1P_2)l\nu_l$ depends on five variables: $P_1P_2$ invariant mass squared $s$, $l\nu_l$ invariant mass squared $q^2$, the angles $\theta$, $\theta^*$ and $\chi$. See the definitions in Fig.~\ref{angle}.
Considering the narrow width approximation
\begin{equation}
\frac{1}{(s-m_{2}^2)^2+m_{2}^2\Gamma_{V}^2}\xrightarrow[]{\Gamma_{V}\to 0}\frac{\pi}{m_{2}\Gamma_{V}}\delta(s-m_{2}^2)
\end{equation}
in the four-body phase space integral, one arrives at
\begin{equation}
\frac{d\Gamma\left(D\rightarrow V(\rightarrow P_1P_2)l^+\nu_l\right)}{dq^2 d\cos\theta d\chi d\cos\theta^*}
=\frac{3 G_F^2 |V_{cq}|^2 v|\vec p_2|}{(2\pi)^4 128m_1^2}Br(V\to P_1P_2)\frac{1}{|\vec p_3|^2}H(\lambda_W, \lambda_W^\prime) L(\lambda_W, \lambda_W^\prime),
\end{equation}
which is then generally expressed as
\begin{equation}
\frac{d\Gamma\left(D\rightarrow V(\rightarrow P_1P_2)l^+\nu_l\right)}{dq^2 d\cos\theta d\frac{\chi}{2\pi}d\cos\theta^*}
=\frac{G_F^2|V_{cq}|^2|\vec{p}_2|q^2v^2}{12(2\pi)^3m_1^2}Br(V\rightarrow P_1P_2)W(\theta,\theta^*,\chi),
\end{equation}
and the angular distribution $W(\theta,\theta^*,\chi)$ is given by
\begin{equation}
\begin{aligned}
W\left(\theta,\theta^*,\chi\right)=&\frac{9}{32}\left(1+\cos^2\theta\right)\sin^2\theta^*\mathcal{H}_U+
\frac{9}{8}\sin^2\theta\cos^2\theta^*\mathcal{H}_L+\frac{9}{16}\cos\theta\sin^2\theta^*\mathcal{H}_P\\
&-\frac{9}{16}\sin^2\theta\sin^2\theta^*\cos2\chi\mathcal{H}_T+\frac{9}{8}\sin\theta\sin2\theta^*\cos\chi\mathcal{H}_A\\
&+\frac{9}{16}\sin2\theta\sin2\theta^*\cos\chi\mathcal{H}_I-\frac{9}{8}\sin\theta\sin2\theta^*\sin\chi\mathcal{H}_{II}\\
&-\frac{9}{16}\sin2\theta\sin2\theta^*\sin\chi\mathcal{H}_{IA}+\frac{9}{16}\sin^2\theta\sin^2\theta^*\sin2\chi\mathcal{H}_{IT}\\
+&\delta_l\left[\frac{9}{4}\cos^2\theta^*\mathcal{H}_S-\frac{9}{2}\cos\theta\cos^2\theta^*\mathcal{H}_{SL}+\frac{9}{4}\cos^2\theta\cos^2\theta^*\mathcal{H}_L\right.\\
&+\frac{9}{16}\sin^2\theta\sin^2\theta^*\mathcal{H}_U+\frac{9}{8}\sin^2\theta\sin^2\theta^*\cos2\chi\mathcal{H}_T\\
&+\frac{9}{4}\sin\theta\sin2\theta^*\cos\chi\mathcal{H}_{ST}-\frac{9}{8}\sin2\theta\sin2\theta^*\cos\chi\mathcal{H}_I\\
&-\frac{9}{4}\sin\theta\sin2\theta^*\sin\chi\mathcal{H}_{ISA}+\frac{9}{8}\sin2\theta\sin2\theta^*\sin\chi\mathcal{H}_{IA}\\
&\left.-\frac{9}{8}\sin^2\theta\sin^2\theta^*\sin2\chi\mathcal{H}_{IT}\right].
\end{aligned}
\label{eqan}
\end{equation}

Note that the observables such as differential decay rate, forward-backward asymmetry, polarization of $\tau$ and vector mesons, can be constructed from three-body decays directly, without the help of Eq.~\eqref{eqan}, which is derived assuming the narrow width limit. The uncertainty due to the narrow width approximation influence only the following $W_i$ measurements. Even so, the finite-width effect is generally very small for vector menson, less than a few percent, as explored in Ref.~\cite{Cheng:2020iwk}.

\section{THE PHYSICAL OBSERVABLES}
\label{Sec:observable}
To study the effect of the lepton mass and provide a more detailed physical picture in semileptonic decays beyond the branching fraction, we can also define other physical observables that can be measured experimentally, such as forward-backward asymmetry $\left(\mathcal{A}_{FB}^l\right)$, longitudinal $\left(P_L^l\right)$ and transverse $\left(P_T^l\right)$ polarization of the charged lepton, longitudinal polarization $\left(F_L^l(V)\right)$ of the final vector meson, leptonic convexity parameter $\left(C_F^l\right)$, and trigonometric momentum $\left(W_i\right)$ in the angular distribution. These observables are expressed again by the above helicity structure functions. The hadronic convexity parameter $\left(C_F^h\right)$ is simply related to the longitudinal polarization of vector, and we will not repetitively calculate $C_F^h$.

First, we consider the forward-backward asymmetry. The ``forward" region requires the $\theta\in\left[0,\pi/2\right]$ and the ``backward" region $\theta\in\left[\pi/2,\pi \right]$. Then, $\mathcal{A}_{FB}^l$ is defined as
\begin{equation}
\begin{aligned}
\mathcal{A}_{FB}^l(q^2)
&=\frac{\int_0^1d\cos\theta \frac{d\Gamma}{dq^2d\cos\theta}-\int_{-1}^0d\cos\theta \frac{d\Gamma}{dq^2d\cos\theta}}{\int_0^1d\cos\theta \frac{d\Gamma}{dq^2d\cos\theta}+\int_{-1}^0d\cos\theta \frac{d\Gamma}{dq^2d\cos\theta}}\\
&=\frac{3}{4}\frac{H_P-4\delta_lH_{SL}}{H_{tot}}.
\end{aligned}
\label{eqafb}
\end{equation}

Next, we consider the polarization observables, which can be derived from the three-body decay. We then define a system where the leptons lie in the $xz$ plane, leading to $k_1=\left(E_1,\; |\vec{k_1}|\sin\theta,\; 0,\; |\vec{k_1}|\cos\theta\right)$. That is, the observables have no $\chi$ dependence and integration over $\chi$ trivially gives $2\pi$ factor. Generally, the expression of the spin four-vector $s$ is \cite{Greiner:1993qp}
\begin{equation}
s^\mu=\left(\frac{\vec{k}_1\cdot\hat{\vec{s}}}{m_l}, \; \hat{\vec{s}}+\frac{\vec{k}_1\left(\vec{k}_1\cdot\hat{\vec{s}}\right)}{m_l\left(k_1^0+m_l\right)}\right),
\end{equation}
which can be obtained by imposing a Lorentz transformation to $(0, \vec s)$, and $\vec s$ is a unit vector defining the direction of spin in the rest frame.
For the longitudinal polarization of the charged lepton, $\vec{k}_1\cdot\hat{\vec{s}}=|\vec{k}_1|$, we obtain the longitudinal polarization vector:
\begin{equation}
s_L^\mu=\frac{1}{m_l}\left(|\vec{k}_1|,\; E_1\sin\theta,\; 0,\; E_1\cos\theta\right),
\end{equation}
which satisfies $s_{L\mu}s_L^\mu=-1$, $s_{L\mu}k_1^{\mu}=0$.

For the leptonic tensor $L_{\mu\nu}$ in Eq.(\ref{eql}), one sums over the spins in the the product of Dirac spinors,
$\sum\limits_s u(p,s)\bar{u}(p,s)=\left(\slashed{p}+m\right)$ and $\sum\limits_s v(p,s)\bar{v}(p,s)=\left(\slashed{p}-m\right)$.
While the lepton is polarized, one has
$u(p,s_L)\bar{u}(p,s_L)=\frac{1}{2}\left[\left(\slashed{p}+m\right)\left(1+\gamma_5\slashed{s}\right)\right]$ and $v(p,s_L)\bar{v}(p,s_L)=\frac{1}{2}\left[\left(\slashed{p}-m\right)\left(1+\gamma_5\slashed{s}\right)\right]$. The expression of the leptonic tensor for the longitudinally polarized lepton is
\begin{equation}
\begin{aligned}
L_{\mu\nu}(s_L)=4&\left(k_{1\mu}k_{2\nu}+k_{1\nu}k_{2\mu}-k_1\cdot k_2g_{\mu\nu}\pm i\varepsilon_{\mu\nu\alpha\beta}k_1^\alpha k_2^\beta\right.\\
&\left.\mp m_ls_{L\mu}k_{2\nu}\mp m_ls_{L\nu}k_{2\mu}\pm m_ls_L\cdot k_2g_{\mu\nu}- im_l\varepsilon_{\mu\nu\alpha\beta}s_L^\alpha k_2^\beta\right)
\end{aligned}
\label{eqltensorfull}
\end{equation}
where the upper/lower sign refers to the $(l^-\bar{\nu}_l)/(l^+\nu_l)$ configuration.
In Ref.~\cite{Gutsche:2015mxa} the leptonic tensor is written as
\begin{equation}
L_{\mu\nu}(s_L)=\mp8m_l\left(s_{L\mu}k_{2\nu}+s_{L\nu}k_{2\mu}-s_L\cdot k_2g_{\mu\nu}\pm i\varepsilon_{\mu\nu\alpha\beta}s_L^\alpha k_2^\beta\right),
\label{eqltensor}
\end{equation}
i.e., making the replacement $k_1^\mu\rightarrow\mp m_ls_L^\mu$ from the unpolarized case. Equations \eqref{eqltensorfull} and \eqref{eqltensor} are equivalent in the sense of $L(\lambda_W, \lambda_W^\prime)$, cf.~Eq.~\eqref{eq:LHhelicity}.

With Eq.~\eqref{eqltensor} we obtain the polarized differential decay distribution:
\begin{equation}
\begin{aligned}
\frac{d\Gamma(s_L)}{dq^2}
&=\frac{G_F^2|V_{cq}|^2|\vec{p}_2|q^2v^2}{12(2\pi)^3m_1^2}\left[-3\delta_l\left|H_t\right|^2+\left(1-\delta_l\right)\left(|H_+|^2+|H_-|^2+|H_0|^2\right)\right]\\
&=\frac{G_F^2|V_{cq}|^2|\vec{p}_2|q^2v^2}{12(2\pi)^3m_1^2}\left[\mathcal{H}_U+\mathcal{H}_L-\delta_l\left(\mathcal{H}_U+\mathcal{H}_L+3\mathcal{H}_S\right)\right].
\end{aligned}
\label{eqsz}
\end{equation}
The longitudinal polarization of the lepton is then defined as the ratio of polarized decay distribution Eq.~\eqref{eqsz} to the unpolarized decay distribution Eq.~\eqref{equn}:
\begin{equation}
\begin{aligned}
P_L^l(q^2)&=\frac{\mathcal{H}_U+\mathcal{H}_L-\delta_l\left(\mathcal{H}_U+\mathcal{H}_L+3\mathcal{H}_S\right)}{\mathcal{H}_{tot}}\\
\end{aligned}
\label{eqpl}
\end{equation}

Similarly, one can obtain the definition of the leptonic transverse polarization $P_T^l(q^2)$, where the leptonic polarization direction is perpendicular to its momentum direction:
\begin{equation}
s_T^\mu
=\left(0, \; \hat{\vec{s}}_T\right)=\left(0,\; \cos\theta,\; 0,\; -\sin\theta\right).
\end{equation}
$\hat{\vec s}_T$ is obtained from rotating $\hat{\vec s}_L=(\sin\theta, 0, \cos\theta)$
by $\pi/2$ in the counterclockwise direction.
Making the substitution $k_1^\mu\rightarrow\mp m_ls_T^\mu$, one obtains the transversely polarized leptonic tensor, and the resulting differential decay distribution is
\begin{equation}
\begin{aligned}
\frac{d\Gamma(s_T)}{dq^2}
&=\frac{G_F^2|V_{cq}|^2|\vec{p}_2|q^2v^2}{12(2\pi)^3m_1^2}\frac{3\pi\sqrt{\delta_l}}{4\sqrt{2}}\left[-\left|H_+\right|^2+\left|H_-\right|^2-2\Re (H_0H_t^\dagger)\right]\\
&=\frac{G_F^2|V_{cq}|^2|\vec{p}_2|q^2v^2}{12(2\pi)^3m_1^2}\frac{3\pi\sqrt{\delta_l}}{4\sqrt{2}}\left[-\mathcal{H}_P-2\mathcal{H}_{SL}\right].
\end{aligned}
\label{eqsx}
\end{equation}
The transverse polarization of lepton will be
\begin{equation}
\begin{aligned}
P_T^l(q^2)&=-\frac{3\pi\sqrt{\delta_l}}{4\sqrt{2}}\frac{\mathcal{H}_P+2\mathcal{H}_{SL}}{\mathcal{H}_{tot}}.\\
\end{aligned}
\label{eqpt}
\end{equation}

In addition, the fourfold decay distribution allows us to define more physical quantities. The angular distribution can be normalized as
\begin{equation}
\widetilde{W}\left(\theta^*, \theta, \chi\right)=\frac{W\left(\theta^*, \theta, \chi\right)}{\mathcal{H}_{tot}}.
\end{equation}
Integrating $W\left(\theta^*, \theta, \chi\right)$ over $\cos\theta^*$ and $\chi$, one can recover the angular distribution of $\theta$ appearing in the $D\to V l\nu_l$ transition (Eq.~\eqref{eq:Gammathree}):
\begin{equation}
\begin{aligned}
W\left(\theta\right)
=&\frac{3}{8}\times\left[\left(1+2\delta_l\right)\mathcal{H}_U+2\mathcal{H}_L+4\delta_l\mathcal{H}_S
+\left(2\mathcal{H}_P-8\delta_l\mathcal{H}_{SL}\right)\cos\theta\right.\\
&\left.+((1-2\delta_l)\mathcal{H}_U-2(1-2\delta_l)\mathcal{H}_L)\cos^2\theta\right].
\end{aligned}
\end{equation}
The normalized angular distribution of $\theta$ is then
\begin{equation}
\widetilde{W}(\theta)=\frac{W(\theta)}{\mathcal{H}_{tot}}
=\frac{a+b\cos\theta+c\cos^2\theta}{2\left(a+c/3\right)},
\end{equation}
where
\begin{equation}
\begin{aligned}
a&=3/8\times\left[\left(1+2\delta_l\right)\mathcal{H}_U+2\mathcal{H}_L+4\delta_l\mathcal{H}_S\right],\\
b&=3/8\times\left(2\mathcal{H}_P-8\delta_l\mathcal{H}_{SL}\right),\\
c&=3/8\times\left[\left(1-2\delta_l\right)\mathcal{H}_U-2\left(1-2\delta_l\right)\mathcal{H}_L\right].
\end{aligned}
\end{equation}
By the above coefficient definition, it is clear that the linear coefficient indicates the forward-backward asymmetry
\begin{equation}
\mathcal{A}_{FB}(q^2)=\frac{b}{2\left(a+c/3\right)}=\frac{3}{4}\frac{\mathcal{H}_P-4\delta_l\mathcal{H}_{SL}}{\mathcal{H}_{tot}}.
\end{equation}
The leptonic convexity parameter is defined as
\begin{equation}
\begin{aligned}
C_F^l(q^2)&=\frac{d^2\widetilde{W}(\theta)}{d(\cos\theta)^2}
=\frac{c}{a+c/3}=\frac{3}{4}\left(1-2\delta_l\right)\frac{\mathcal{H}_U-2\mathcal{H}_L}{\mathcal{H}_{tot}}.\\
\end{aligned}
\end{equation}

In the same manner, we can obtain the $\theta^*$ angular distribution by integrating the $W(\theta,\theta^*,\chi)$ over $\theta$ and $\chi$:
\begin{equation}
\begin{aligned}
W(\theta^*)
=\frac{3}{4}\times\left[\left(1+\delta_l\right)\mathcal{H}_U+\left(-\left(1+\delta_l\right)\mathcal{H}_U
+2\left(1+\delta_l\right)\mathcal{H}_L+6\delta_l\mathcal{H}_S\right)\cos^2\theta^*\right].
\end{aligned}
\end{equation}
The normalized $\theta^*$ angular distribution is described by
\begin{equation}
\widetilde{W}(\theta^*)=\frac{W(\theta^*)}{\mathcal{H}_{tot}}=\frac{a'+c'\cos^2\theta^*}{2a'+2/3c'},
\end{equation}
where
\begin{equation}
\begin{aligned}
a'&=\left(1+\delta_l\right)\mathcal{H}_U\\
c'&=-\left(1+\delta_l\right)\mathcal{H}_U+2\left(1+\delta_l\right)\mathcal{H}_L+6\delta_l\mathcal{H}_S.
\end{aligned}
\end{equation}
The hadronic convexity parameter can be extracted by taking the second derivative of $\widetilde{W}(\theta^*)$ as
\begin{equation}
\begin{aligned}
C^h_F(q^2)&=\frac{d^2\widetilde{W}(\theta^*)}{d(\cos\theta^*)^2}
=\frac{c'}{a'+c'/3}=-\frac{3}{2}\frac{\mathcal{H}_U-2\mathcal{H}_L+
\delta_l\left(\mathcal{H}_U-2\mathcal{H}_L-6\mathcal{H}_S\right)}{\mathcal{H}_{tot}}.\\
\end{aligned}
\end{equation}
The longitudinal polarization fraction of the final vector meson is given by
\begin{equation}
\begin{aligned}
F_L^l(q^2)&=\frac{d\Gamma(\lambda_V=0)/dq^2}{d\Gamma/dq^2}=\frac{(1+\delta_l)\mathcal{H}_L+3\delta_l\mathcal{H}_S}{\mathcal{H}_{tot}},\\
\end{aligned}
\label{eqfl}
\end{equation}
and the transverse polarization fraction is $F_T^l(q^2)=1-F_L^l(q^2)$. The hadronic convexity parameter is related to the longitudinal polarization of vector meson by
\begin{equation}
C_F^h(q^2)=\frac{3}{2}(3F_L(q^2)-1).
\end{equation}

Moreover, to examine more helicity structure functions in Eq.~\eqref{eqan}, we can define the trigonometric moments through the normalized angular decay distribution $\widetilde{W}(\theta^*, \theta, \chi)$
\begin{equation}
W_i=\int d\cos\theta d\cos\theta^*d\left(\chi/2\pi\right)M_i\left(\theta,\theta^*,\chi\right)\widetilde{W}\left(\theta,\theta^*,\chi\right)=\left\langle M_i\left(\theta,\theta^*,\chi\right)\right\rangle.
\end{equation}
Some explicit examples include
\begin{equation}
\begin{aligned}
W_T(q^2)&=\left\langle\cos2\chi\right\rangle=-\frac{1}{2}\left(1-2\delta_l\right)\frac{\mathcal{H}_T}{\mathcal{H}_{tot}},\\
W_I(q^2)&=\left\langle\cos\theta\cos\theta^*\cos\chi\right\rangle=\frac{9\pi^2\left(1-2\delta_l\right)}{512}\frac{\mathcal{H}_I}{\mathcal{H}_{tot}},\\
W_A(q^2)&=\left\langle\sin\theta\sin\theta^*\cos\chi\right\rangle=\frac{3\pi}{16}\frac{\mathcal{H}_A+2\delta_l\mathcal{H}_{ST}}{\mathcal{H}_{tot}}.\\
\end{aligned}
\label{eqw}
\end{equation}
For final mesons being pseudoscalars, the helicity amplitudes $H_{\pm}=0$, which leads to vanishing trigonometric moments. Therefore, Eq.~\eqref{eqw} is only applied to $D\rightarrow V$ transitions.

In practice, due to the limited statistics in the experiment, we show the average values of all mentioned physical observables. In such a case, we reinstate the phase factor $C(q^2)=\left|\vec{p_2}\right|(q^2-m_l^2)^2/q^2$ in both numerator and denominator and integrate them separately. More explicitly,
\begin{equation}
\langle \mathcal{A}_{FB}^l\rangle=\frac{3}{4}\frac{\mathlarger{\int} dq^2C(q^2)\left(\mathcal{H}_P-4\delta_l\mathcal{H}_{SL}\right)}{\mathlarger{\int} dq^2C(q^2)\left[\mathcal{H}_U+\mathcal{H}_L+\delta_l\left(\mathcal{H}_U+\mathcal{H}_L+3\mathcal{H}_S\right)\right]}.
\end{equation}
A similar operation holds for all others. We also present the full momentum dependence of $\mathcal{A}_{FB}^l$, $P_L^l$ and $P_T^l$.

For the case in which the charge of the lepton is negative, i.e., $l^-\bar{\nu}_l$, the relative sign in Eq.~\eqref{eql} and Eq.~\eqref{eqltensor} will influence the expressions of some observables. To better compare with experimental and other theoretical results, we also give the definitions of forward-backward asymmetry, leptonic longitudinal and transverse polarization in such a case:
\begin{equation}
\begin{aligned}
&\mathcal{A}_{FB}(q^2)=-\frac{3}{4}\frac{\mathcal{H}_P+4\delta_l\mathcal{H}_{SL}}{\mathcal{H}_{tot}},\\
&P_L^l(q^2)=-\frac{\mathcal{H}_U+\mathcal{H}_L-\delta_l\left(\mathcal{H}_U+\mathcal{H}_L+3\mathcal{H}_S\right)}{\mathcal{H}_{tot}},\\
&P_T^l(q^2)=-\frac{3\pi\sqrt{\delta_l}}{4\sqrt{2}}\frac{\mathcal{H}_P-2\mathcal{H}_{SL}}{\mathcal{H}_{tot}}.\\
\end{aligned}
\label{eqp}
\end{equation}
The expressions for other observables do not change.

\section{NUMERICAL RESULTS AND DISCUSSION}
\label{Sec:results}

In this section, we will calculate the branching fractions of $D_{(s)}$ and $B_{(s)}$ to $(P,V)l^+\nu_l$ and the average values of $\mathcal{A}_{FB}^l$, $P_L^l$, $P_T^l$, $C_F^l$, and $W_i$, on the basis of related form factors and the abovementioned helicity amplitudes. The average values for these observables and the $q^2$ dependence are presented in the tables and figures below.

Concerning the related form factors in these physical quantities equations, we stick to the predictions in the covariant light-front quark model (CLFQM) \cite{Cheng:2003sm,Verma:2011yw}.
The momentum dependence of form factors can be parameterized as \footnote{The shape of form factor is generally smooth, and such a parametrization
works very accurately. The difference between it and the exact numerical value is less than 3\%. However, there are indeed exceptions. See discussions between Eqs.~(33) and (34) in Ref.~\cite{Cheng:2003sm}. For the $B(D)\to {}^1P_1$ decays, a new parametrisation is proposed, but it is not relevant to our current study. }
\begin{equation}
F(q^2)=\frac{F(0)}{1-a\left(q^2/m_{D}^2\right)+b\left(q^2/m_{D}^2\right)^2}.
\end{equation}
The values of relevant parameters that appear in form factors are listed in Table $\mathrm{\ref{tabff}}$.

For the final state of the $D_{(s)}$ transitions, the pseudoscalar mesons $P$ contain $\eta, \eta', \pi^0, K^0$, $\bar{K^0}$ and the vector mesons $V$ contain $\rho, \omega, \phi, K^*$,$\bar{K^*}$; for the $B_{(s)}$ transitions, the pseudoscalar mesons $P$ contain $\eta, \eta', \pi^0, \bar{D}^0, D_s^-$ and the vector mesons $V$ contain $\rho, \omega, K^{*-}, \bar{D}^{*0}$, $D_s^{*-}$. For  $\eta$ and $\eta'$, we need to consider the quark mixing scheme \cite{Feldmann:1998vh,Feldmann:1999uf}
\begin{equation}
\left( \begin{array}{c}
\eta\\
\eta'
\end{array} \right)=
\left( \begin{array}{cc}
\cos\phi&-\sin\phi\\
\sin\phi&\cos\phi
\end{array} \right)
\left( \begin{array}{c}
\eta_q\\
\eta_s
\end{array} \right),
\end{equation}
where $\eta_q=\frac{1}{\sqrt{2}}(\mu\bar{\mu}+d\bar{d})$, $\eta_s=s\bar{s}$, and the mixing angle $\phi=39.3^{\circ}$ from Ref.~\cite{Zyla:2020zbs,Dai:2017tew}.

\begin{table}[!htbp]
\centering
\caption{The form factor parameters predicted by CLFQM in $D_{(s)}/B_{(s)}\rightarrow P(V)$ transitions. The values are taken from Ref.~\cite{Verma:2011yw}.}
\begin{tabular}{cccc|cccc|cccc|cccc}
\Xhline{1pt}
$F$&$F(0)$&$a$&$b$&$F$&$F(0)$&$a$&$b$&$F$&$F(0)$&$a$&$b$&$F$&$F(0)$&$a$&$b$\\
\Xhline{1pt}
$F_1^{D\pi}$&0.66&1.19&0.35&$F_0^{D\pi}$&0.66&0.51&0.00&$F_1^{B\pi}$&0.25&1.70&0.90&$F_0^{B\pi}$&0.25&0.82&0.09\\
$F_1^{D\eta_q}$&0.71&1.13&0.27&$F_0^{D\eta_q}$&0.71&0.43&$-0.01$&$F_1^{B\eta_q}$&0.29&1.63&0.74&$F_0^{B\eta_q}$&0.29&0.75&0.04\\
$F_1^{DK}$&0.79&1.05&0.25&$F_0^{DK}$&0.79&0.47&$-0.00$&$F_1^{BD}$&0.67&1.22&0.36&$F_0^{BD}$&0.67&0.63&$-0.01$\\
\hline
$F_1^{D_sK}$&0.66&1.11&0.48&$F_0^{D_sK}$&0.66&0.56&0.04&$F_1^{B_sK}$&0.23&1.88&1.58&$F_0^{B_sK}$&0.23&1.05&0.35\\
$F_1^{D_s\eta_s}$&0.76&1.02&0.40&$F_0^{D_s\eta_s}$&0.76&0.60&0.04&$F_1^{B_sD_s}$&0.67&1.28&0.52&$F_0^{B_sD_s}$&0.67&0.69&$0.07$\\
\Xhline{1pt}
$V^{D\rho}$&0.88&1.23&0.40&$A_0^{D\rho}$&0.69&1.08&0.45&$A_1^{D\rho}$&0.60&0.46&0.01&$A_2^{D\rho}$&0.47&0.89&0.23\\
$V^{D\omega}$&0.85&1.24&0.45&$A_0^{D\omega}$&0.64&1.08&1.50&$A_1^{D\omega}$&0.58&0.49&0.02&$A_2^{D\omega}$&0.49&0.95&0.28\\
$V^{DK^*}$&0.98&1.10&0.32&$A_0^{DK^*}$&0.78&1.01&0.34&$A_1^{DK^*}$&0.72&0.45&0.01&$A_2^{DK^*}$&0.60&0.89&0.21\\
\hline
$V^{D_sK^*}$&0.87&1.13&0.69&$A_0^{D_sK^*}$&0.61&0.90&0.62&$A_1^{D_sK^*}$&0.56&0.59&0.08&$A_2^{D_sK^*}$&0.46&0.90&0.43\\
$V^{D_s\phi}$&0.98&1.04&0.54&$A_0^{D_s\phi}$&0.72&0.92&0.62&$A_1^{D_s\phi}$&0.69&0.56&0.07&$A_2^{D_s\phi}$&0.59&0.90&0.38\\
\hline
$V^{B\rho}$&0.29&1.77&1.06&$A_0^{B\rho}$&0.32&1.67&1.01&$A_1^{B\rho}$&0.24&0.86&0.15&$A_2^{B\rho}$&0.22&1.56&0.85\\
$V^{B\omega}$&0.27&1.81&1.18&$A_0^{B\omega}$&0.28&1.62&1.22&$A_1^{B\omega}$&0.23&0.91&0.18&$A_2^{B\omega}$&0.21&1.62&0.97\\
$V^{BD^*}$&0.77&1.25&0.38&$A_0^{BD^*}$&0.68&1.21&0.36&$A_1^{BD^*}$&0.65&0.60&0.00&$A_2^{BD^*}$&0.61&1.12&0.31\\
\hline
$V^{B_sK^*}$&0.23&2.03&2.27&$A_0^{B_sK^*}$&0.25&1.95&2.20&$A_1^{B_sK^*}$&0.19&1.24&0.62&$A_2^{B_sK^*}$&0.16&1.83&1.85\\
$V^{B_sD_s^*}$&0.75&1.37&0.67&$A_0^{B_sD_s^*}$&0.66&1.33&0.63&$A_1^{B_sD_s^*}$&0.62&0.76&0.13&$A_2^{B_sD_s^*}$&0.57&1.25&0.56\\
\Xhline{1pt}
\end{tabular}
\label{tabff}
\end{table}

\begin{table}
\caption{Branching fraction for the semileptonic decays of $D^+$ and $D_s^+$, compared to the PDG~\cite{Zyla:2020zbs} and BES${\mathrm{\uppercase\expandafter{\romannumeral3}}}$ results. All values are in unit of $10^{-3}$}
\begin{adjustbox}{center}
\begin{tabular}{!{\vrule width 1pt}c!{\vrule width 1pt}c|ccc|ccc!{\vrule width 1pt}}
\Xhline{1pt}
\multirow{2}*{\;}&
\;&\multicolumn{3}{c|}{$e$ mode}&\multicolumn{3}{c!{\vrule width 1pt}}{$\mu$ mode}\\
\;&\;&CLFQM&PDG&BES${\mathrm{\uppercase\expandafter{\romannumeral3}}}$&CLFQM&PDG&BES${\mathrm{\uppercase\expandafter{\romannumeral3}}}$\\
\Xhline{1pt}
\multirow{4}*{$D^+\rightarrow P$}&
$D^+\rightarrow\eta l^+\nu_l$
&$1.20$&$1.11\pm0.07$&\;
&$1.16$&\;&$1.04\pm0.15$ \cite{BESIII:2020dtz}\\

&$D^+\rightarrow\eta' l^+\nu_l$
&$0.179$&$0.20\pm0.04$&\;&$0.169$&\;&\;\\

&$D^+\rightarrow\pi^0 l^+\nu_l$&$4.09$&$3.72\pm0.17$&$3.63\pm0.13$ \cite{Ablikim:2017lks}
&$4.04$&$3.50\pm0.15$&\;\\

&$D^+\rightarrow\bar{K}^0l^+\nu_l$
&$103.2$&$87.3\pm1.0$&$86.0\pm2.1$ \cite{Ablikim:2017lks}
&$100.7$&$87.6\pm1.9$&\;\\
\hline
\multirow{3}*{$D^+\rightarrow V$}&$D^+\rightarrow\rho l^+\nu_l$&$2.32$&$2.18^{+0.17}_{-0.25}$&\;&$2.22$&$2.4\pm0.4$&\;\\

&$D^+\rightarrow\omega l^+\nu_l$&$2.07$&$1.69\pm0.11$&$1.69\pm0.11$ \cite{BESIII:2020dbj}
&$1.98$&\;&$1.77\pm0.29$ \cite{BESIII:2020dbj}\\

&$D^+\rightarrow\bar{K}^{*0}l^+\nu_l$&$73.2$&$54.0\pm1.0$&\;
&$69.3$&$52.7\pm1.5$&\;\\
\hline
\multirow{3}*{$D_s^+\rightarrow P$}&$D_s^+\rightarrow\eta l^+\nu_l$
&$21.9$&$23.2\pm0.8$&$23.0\pm3.9$ \cite{Ablikim:2017omq}&
$21.5$&$24\pm5.0$&$24.2\pm5.7$ \cite{Ablikim:2017omq}\\

&$D_s^+\rightarrow\eta' l^+\nu_l$&$8.82$&$8.0\pm0.7$&$9.3\pm3.5$ \cite{Ablikim:2017omq}
&$8.41$&$11\pm5.0$&$10.6\pm6.1$ \cite{Ablikim:2017omq}\\

&$D_s^+\rightarrow K^0 l^+\nu_l$&$2.54$&$3.4\pm0.4$&\;
&$2.49$&\;&\;\\
\hline
\multirow{2}*{$D_s^+\rightarrow V$}&$D_s^+\rightarrow\phi l^+\nu_l$&$30.7$&$23.9\pm1.6$&$22.6\pm5.4$ \cite{Ablikim:2017omq}
&$28.9$&$19.0\pm5.0$&$19.4\pm6.2$ \cite{Ablikim:2017omq}\\

&$D_s^+\rightarrow K^{*0} l^+\nu_l$&$1.90$&$2.15\pm0.28$&\;
&$1.82$&\;&\;\\
\Xhline{1pt}
\end{tabular}
\end{adjustbox}
\label{tabbrd}
\end{table}

\begin{table}[!htbp]
\centering
\caption{Branching fraction for the semileptonic decays of $B^+$ and $B_s$, compared to the PDG \cite{Tanabashi:2018oca} results.}
\begin{tabular}{!{\vrule width 1pt}c!{\vrule width 1pt}c|cccc!{\vrule width 1pt}}
\Xhline{1pt}
\;&\;&$e$ mode&PDG ($l^+\nu_l$)&$\tau$ mode&PDG ($\tau^+\nu_{\tau}$)\\
\Xhline{1pt}
\multirow{4}*{$B^+\rightarrow P$}&
$B^+\rightarrow\eta l^+\nu_l$&$4.96\times10^{-5}$&$(3.9\pm0.5)\times10^{-5}$&$3.03\times10^{-5}$&\;\\

&$B^+\rightarrow\eta' l^+\nu_l$&$2.41\times10^{-5}$&$(2.3\pm0.8)\times10^{-5}$&$1.28\times10^{-5}$&\;\\

&$B^+\rightarrow\pi^0 l^+\nu_l$&$7.20\times10^{-5}$&$(7.8\pm0.27)\times10^{-5}$&$4.89\times10^{-5}$&\;\\

&$B^+\rightarrow\bar D^0 l^+\nu_l$&$2.59\times10^{-2}$&$(2.35\pm0.09)\times10^{-2}$
&$0.78\times10^{-2}$&$(0.77\pm0.25)\times10^{-2}$\\
\hline
\multirow{3}*{$B^+\rightarrow V$}&
$B^+\rightarrow\rho l^+\nu_l$&$2.00\times10^{-4}$&$(1.58\pm0.11)\times10^{-4}$&$1.09\times10^{-4}$&\;\\

&$B^+\rightarrow\omega l^+\nu_l$&$1.89\times10^{-4}$&$(1.19\pm0.09)\times10^{-4}$&$1.00\times10^{-4}$&\;\\

&$B^+\rightarrow\bar D^{*0} l^+\nu_l$&$6.67\times10^{-2}$&$(5.66\pm0.22)\times10^{-2}$
&$1.66\times10^{-2}$&$(1.88\pm0.20)\times10^{-2}$\\
\hline
\multirow{2}*{$B_s\rightarrow P$}&
$B_s\rightarrow K^- l^+\nu_l$&$9.23\times10^{-5}$&\;&$6.18\times10^{-5}$&\;\\

&$B_s\rightarrow D_s^- l^+\nu_l$&$2.41\times10^{-2}$&$\;$&$0.72\times10^{-2}$&\;\\
\hline
\multirow{2}*{$B_s\rightarrow V$}
&$B_s\rightarrow K^{*-} l^+\nu_l$&$3.01\times10^{-4}$&\;&$1.56\times10^{-4}$&\;\\

&$B_s\rightarrow D_s^{*-} l^+\nu_l$&$5.91\times10^{-2}$&\;&$1.46\times10^{-2}$&\;\\
\Xhline{1pt}
\end{tabular}
\label{tabbrb}
\end{table}

For the leptons involved, we consider the electron and muon mode for the $D_{(s)}$ transition, separately. In addition, the corresponding tau mode is prohibited by the phase space. By contrast, for the $B_{(s)}$ transition, the electron and tau mode are included, where the results of the electron and muon mode are almost identical, as expected, due to the large mass of $B_{(s)}$, and the difference between them is beyond the current experimental uncertainties.

\begin{table}[!htbp]
\centering
\caption{Forward-backward asymmetry, lepton polarization, leptonic convexity parameter, and longitudinal polarization of the vector meson for semileptonic decays of $D^+$ and $D_s^+$, compared with other predictions from the CCQM~\cite{Ivanov:2019nqd} and RQM~\cite{Faustov:2019mqr}. The dash indicates that the value does not exist. In Ref.~\cite{Faustov:2019mqr}, given the specific scale, some numbers of $\langle A_{FB}^e\rangle$ and $\langle P_T^e\rangle$ (with orders of $10^{-6}$ and $10^{-3}$, respectively) are shown as 0.}
\begin{tabular}{!{\vrule width 1pt}c!{\vrule width 1pt}cccccccccccc!{\vrule width 1pt}}
\Xhline{1pt}
\;&\;&\;&$\langle\mathcal{A}_{FB}^e\rangle$&$\langle\mathcal{A}_{FB}^\mu\rangle$&$\langle P_L^e\rangle$&$\langle P_L^\mu\rangle$&$\langle P_T^e\rangle(10^{-2})$&$\langle P_T^\mu\rangle$&$\langle C_F^e\rangle$&$\langle C_F^\mu\rangle$&$\langle F_L^e\rangle$&$\langle F_L^\mu\rangle$\\
\Xhline{1pt}
\multirow{12}*{$D^+\rightarrow P$}&
$D^+\rightarrow\eta l^+\nu_l$&CLFQM&$-6.0\times10^{-6}$&$-0.05$&$1.00$&$0.84$&$-0.27$&$-0.43$&$-1.50$&$-1.36$&-&-\\
&\;&CCQM&$-6.4\times10^{-6}$&$-0.06$&$1.00$&$0.83$&$-0.28$&$-0.44$&$-1.50$&$-1.32$&-&-\\
&\;&RQM&\;&$-0.052$ &1.00 &0.85&\;&$-0.40$ &$-1.50$ &$-1.34$&-&-\\

&$D^+\rightarrow\eta' l^+\nu_l$&CLFQM&$-13.2\times10^{-6}$&$-0.10$&$1.00$&$0.71$&$-0.41$&$-0.57$&$-1.50$&$-1.27$&-&-\\
&\;&CCQM&$-13.0\times10^{-6}$&$-0.10$&$1.00$&$0.70$&$-0.42$&$-0.59$&$-1.50$&$-1.19$&-&-\\
&\;&RQM&\;&$-0.097$&1.00 &0.72&\;&$-0.56$ &$-1.50$ &$-1.20$&-&-\\

&$D^+\rightarrow\pi^0 l^+\nu_l$&CLFQM&$-3.4\times10^{-6}$&$-0.04$&$1.00$&$0.90$&$-0.20$&$-0.34$&$-1.50$&$-1.40$&-&-\\
&\;&CCQM&$-4.1\times10^{-6}$&$-0.04$&1.00&0.88&$-0.22$&$-0.36$&$-1.50$&$-1.37$&-&-\\
&\;&RQM&\;&$-0.040$&1.00 &0.89&\;&$-0.36$ &$-1.50$ &$-1.38$&-&-\\

&$D^+\rightarrow\bar K^0 l^+\nu_l$&CLFQM&$-5.8\times10^{-6}$&$-0.05$&$1.00$&$0.84$&$-0.27$&$-0.42$&$-1.50$&$-1.36$&-&-\\
&\;&CCQM&$-6.4\times10^{-6}$&$-0.06$&1.00&0.83&$-0.28$&$-0.43$&$-1.50$&$-1.32$&-&-\\
&\;&RQM&\;&$-0.053$ &1.00 &0.85&\;&$-0.42$ &$-1.50$ &$-1.34$&-&-\\
\hline
\multirow{9}*{$D^+\rightarrow V$}&
$D^+\rightarrow\rho l^+\nu_l$&CLFQM&$-0.24$&$-0.26$&$1.00$&$0.92$&$-0.10$&$-0.13$&$-0.48$&$-0.40$&0.55&0.54\\
&\;&CCQM&$-0.21$&$-0.24$&$1.00$&$0.92$&$-0.09$&$-0.13$&$-0.44$&$-0.36$&0.53&0.51\\
&\;&RQM&$-0.26$&$-0.28$&$1.00$&$0.92$&$\;$&$-0.12$&$-0.42$&$-0.34$&0.52&0.52\\

&$D^+\rightarrow\omega l^+\nu_l$&CLFQM&$-0.24$&$-0.26$&$1.00$&$0.92$&$-0.09$&$-0.12$&$-0.45$&$-0.37$&0.53&0.53\\
&\;&CCQM&$-0.21$&$-0.24$&$1.00$&$0.92$&$-0.09$&$-0.12$&$-0.43$&$-0.35$&0.52&0.50\\
&\;&RQM&$-0.25$&$-0.27$&$1.00$&$0.93$&$\;$&$-0.11$&$-0.39$&$-0.32$&0.51&0.50\\

&$D^+\rightarrow\bar K^{*0} l^+\nu_l$&CLFQM&$-0.19$&$-0.22$&$1.00$&$0.90$&$-0.11$&$-0.15$&$-0.48$&$-0.39$&0.55&0.54\\
&\;&CCQM&$-0.18$&$-0.21$&1.00&0.91&$-0.11$&$-0.15$&$-0.47$&$-0.37$&0.54&0.52\\
&\;&RQM&$-0.22$&$-0.25$&$1.00$&$0.90$&$\;$&$-0.15$&$-0.47$&$-0.37$&0.54&0.54\\
\hline
\multirow{9}*{$D_s^+\rightarrow P$}&
$D_s^+\rightarrow\eta l^+\nu_l$&CLFQM&$-5.6\times10^{-6}$&$-0.05$&$1.00$&$0.84$&$-0.27$&$-0.43$&$-1.50$&$-1.33$&-&-\\
&\;&CCQM&$-6.0\times10^{-6}$&$-0.06$&$1.00$&$0.84$&$-0.27$&$-0.42$&$-1.50$&$-1.33$&-&-\\
&\;&RQM&\;&$-0.043$&\;&0.88&\;&$-0.35$&\;&$-1.37$&-&-\\

&$D_s^+\rightarrow\eta' l^+\nu_l$&CLFQM&$-11.1\times10^{-6}$&$-0.09$&$1.00$&$0.74$&$-0.38$&$-0.55$&$-1.50$&$-1.23$&-&-\\
&\;&CCQM&$-11.2\times10^{-6}$&$-0.09$&$1.00$&$0.75$&$-0.38$&$-0.54$&$-1.50$&$-1.23$&-&-\\
&\;&RQM&\;&$-0.080$&\;&0.77&\;&$-0.51$&\;&$-1.26$&-&-\\

&$D_s^+\rightarrow K^0 l^+\nu_l$&CLFQM&$-5.1\times10^{-6}$&$-0.05$&$1.00$&$0.86$&$-0.25$&$-0.41$&$-1.50$&$-1.35$&-&-\\
&\;&CCQM&$-5.0\times10^{-6}$&$-0.05$&$1.00$&$0.86$&$-0.24$&$-0.39$&$-1.50$&$-1.35$&-&-\\
&\;&RQM&\;&$-0.038$&\;&0.89&\;&$-0.34$&\;&$-1.38$&-&-\\
\hline
\multirow{6}*{$D_s^+\rightarrow V$}
&$D_s^+\rightarrow\phi l^+\nu_l$&CLFQM&$-0.18$&$-0.21$&$1.00$&$0.91$&$-0.11$&$-0.14$&$-0.48$&$-0.38$&0.54&0.53\\
&\;&CCQM&$-0.18$&$-0.21$&$1.00$&$0.91$&$-0.11$&$-0.14$&$-0.43$&$-0.34$&0.53&0.50\\
&\;&RQM&$-0.21$&$-0.24$&$1.00$&$0.90$&$\;$&$-0.15$&$-0.49$&$-0.35$&0.54&0.54\\

&$D_s^+\rightarrow K^{*0} l^+\nu_l$&CLFQM&$-0.22$&$-0.25$&$1.00$&$0.92$&$-0.09$&$-0.12$&$-0.47$&$-0.38$&0.54&0.54\\
&\;&CCQM&$-0.22$&$-0.25$&$1.00$&$0.92$&$-0.09$&$-0.11$&$-0.40$&$-0.33$&0.51&0.49\\
&\;&RQM&$-0.26$&$-0.29$&$1.00$&$0.92$&$\;$&$-0.11$&$-0.41$&$-0.33$&0.52&0.51\\
\Xhline{1pt}
\end{tabular}
\label{tabod}
\end{table}

In Tables~\ref{tabbrd} and~\ref{tabbrb}, we list the numerical results of the branching fractions for $D_{(s)}$ and $B_{(s)}$ decays, respectively, calculated through the helicity formalism in Eq.~\eqref{equn}. The results are in perfect agreement with the values of Ref.~\cite{Cheng:2017pcq}, which are calculated directly via form factors. These two ways are cross-checked well. We also compare our predictions with the results from the PDG~\cite{Tanabashi:2018oca} and BES${\mathrm{\uppercase\expandafter{\romannumeral3}}}$ collaboration \cite{BESIII:2020dtz,BESIII:2020dbj,Ablikim:2017omq,Ablikim:2017lks}, and we find a good agreement among them within the uncertainty. But
they differ a little larger for $D\to K l \nu_l$, and especially for $D\to K^* l\nu_l$. Assuming 15\% uncertainty for the decay rate, our result $\mathcal{B}(D^+\to \bar K^{*0} e^+\nu_e)=73.2\times 10^{-3}$
differs from the experimental one $54.0\times 10^{-3}$ by $1.7 \sigma$, while $2.5 \sigma$ for assuming 10\% uncertainty.
This also happens for CCQM, see Table V in~\cite{Faustov:2019mqr}, where the RQM result agrees better with PDG. It shows again that investigation in a different model is necessary.
For the values of the $B^+\rightarrow P(V)l^+\nu_l$ decay from PDG in Table~\ref{tabbrb}, the $l$ indicates an electron or a muon, not a sum over these two modes.
We also note that in Ref.~\cite{Ivanov:2019nqd}, the branching fractions for $D^+\to D^0 e^+ \nu_e$ and $D_s^+\to D^0 e^+ \nu_e$ are calculated to be on the
order of $10^{-13}$ and $10^{-8}$, respectively. These numbers are far beyond the current and even future scope of the experiment. We will refrain from considering these two channels in our work.

\begin{table}[!htbp]
\centering
\caption{Forward-backward asymmetry, lepton polarization, and leptonic convexity parameter and the longitudinal polarization of vector meson for semileptonic decays of $B^+$ and $B_s$. The dash indicates that the value does not exist. }
\begin{tabular}{!{\vrule width1pt}c!{\vrule width1pt}ccccccccccc!{\vrule width1pt}}
\Xhline{1pt}
\;&\;&$\langle\mathcal{A}_{FB}^e\rangle$&$\langle\mathcal{A}_{FB}^\tau\rangle$&$\langle P_L^e\rangle$&$\langle P_L^\tau\rangle$&$\langle P_T^e\rangle$&$\langle P_T^\tau\rangle$&$\langle C_F^e\rangle$&$\langle C_F^\tau\rangle$&$\langle F_L^e\rangle$&$\langle F_L^{\tau}\rangle$\\
\Xhline{1pt}
\multirow{5}*{$B^+\rightarrow P$}&
$B^+\rightarrow\eta l^+\nu_l$&$-0.39\times10^{-6}$&$-0.29$&$1.00$&$0.11$&$-0.64\times10^{-3}$&$-0.86$&$-1.50$&$-0.60$&-&-\\
&$B^+\rightarrow\eta' l^+\nu_l$&$-0.49\times10^{-6}$&$-0.31$&$1.00$&$0.026$&$-0.72\times10^{-3}$&$-0.87$&$-1.50$&$-0.52$&-&-\\

&$B^+\rightarrow\pi^0 l^+\nu_l$&$-0.35\times10^{-6}$&$-0.28$&$1.00$&$0.087$&$-0.62\times10^{-3}$&$-0.85$&$-1.50$&$-0.59$&-&-\\
&$B^+\rightarrow\bar D^0 l^+\nu_l$&$-1.04\times10^{-6}$&$-0.36$&$1.00$&$-0.32$&$-1.07\times10^{-3}$&$-0.84$&$-1.50$&$-0.27$&-&-\\
\hline
\multirow{3}*{$B^+\rightarrow V$}&
$B^+\rightarrow\rho l^+\nu_l$&$-0.32$&$-0.39$&$1.00$&$0.60$&$-0.18\times10^{-3}$&$-0.10$&$-0.39$&$-0.12$&0.51&0.49\\
&$B^+\rightarrow\omega l^+\nu_l$&$-0.30$&$-0.36$&$1.00$&$0.65$&$-0.15\times10^{-3}$&$-0.06$&$-0.42$&$-0.15$&0.51&0.49\\
&$B^+\rightarrow\bar D^{*0} l^+\nu_l$&$-0.22$&$-0.30$&$1.00$&$0.51$&$-0.29\times10^{-3}$&$-0.10$&$-0.42$&$-0.056$&0.52&0.45\\
\hline
\multirow{2}*{$B_s\rightarrow P$}
&$B_s\rightarrow K^- l^+\nu_l$&$-0.43\times10^{-6}$&$-0.29$&$1.00$&$-0.10$&$-0.72\times10^{-3}$&$-0.86$&$-1.50$&$-0.46$&-&-\\
&$B_s\rightarrow D_s^- l^+\nu_l$&$-1.05\times10^{-6}$&$-0.36$&$1.00$&$-0.33$&$-1.07\times10^{-3}$&$-0.84$&$-1.50$&$-0.26$&-&-\\
\hline
\multirow{2}*{$B_s\rightarrow V$}
&$B_s\rightarrow K^{*-} l^+\nu_l$&$-0.21$&$-0.28$&$1.00$&$0.65$&$-0.17\times10^{-3}$&$-0.13$&$-0.59$&$-0.26$&0.59&0.56\\
&$B_s\rightarrow D_s^{*-} l^+\nu_l$&$-0.22$&$-0.29$&$1.00$&$0.51$&$-0.29\times10^{-3}$&$-0.10$&$-0.4$3&$-0.058$&0.52&0.45\\
\Xhline{1pt}
\end{tabular}
\label{tabob}
\end{table}

In Tables $\mathrm{\ref{tabod}}$ and $\mathrm{\ref{tabob}}$, we list the average values of other observables for the $D_{(s)}$ and $B_{(s)}$ transitions, including the forward-backward asymmetry $\langle \mathcal{A}_{FB}^l\rangle$, leptonic longitudinal and transverse polarization $\langle P_L^l\rangle$, $\langle P_T^l\rangle$, leptonic convexity parameter $\langle C_F^l\rangle$ and longitudinal polarization $\langle F_L^l\rangle$ of the final vector meson. In Table $\mathrm{\ref{tabod}}$, we compare our results with other theoretical predictions from Ref.~\cite{Ivanov:2019nqd} and Ref.~\cite{Faustov:2019mqr}. In these references, the form factors are obtained based on the covariant confining quark model (CCQM) and the relativistic quark model (RQM), separately, and the magnitude of these numerical results agree well.
From these two tables, we find that the average values of the forward-backward asymmetry $A_{FB}^{\mu(\tau)}$ are similar to those for $\mathcal{A}_{FB}^{e}$ for both the $D_{s}\rightarrow V$ and $B_{s}\rightarrow V$ transition, while for the $D_{s}\rightarrow P$ and $B_{s}\rightarrow P$ transitions, we obtain $\langle \mathcal{A}_{FB}^\mu\rangle/\langle \mathcal{A}_{FB}^e\rangle\sim10^4$ and $\langle \mathcal{A}_{FB}^\tau\rangle/\langle \mathcal{A}_{FB}^e\rangle\sim10^7$. Therefore, the lepton mass effect is more apparent for the mode of the pseudoscalar in final state. In fact, it is a natural result of Eq.~\eqref{eqafb}, where for the pseudoscalar case $H_{\pm}=0$, then $\mathcal{H}_P=0$ and $\mathcal{A}_{FB}$ is proportional to the lepton mass squared; the above ratio just corresponds to $m_{\mu}^2/m_{e}^2$ and $m_{\tau}^2/m_{e}^2$, respectively. Concerning the average values of the leptonic transverse polarization, the absolute values of $\langle P_T^{\mu}\rangle$ are much larger than $\langle P_T^{e}\rangle$ for $D_{(s)}$ decay and $|\langle P_T^{\tau}\rangle| \gg |\langle P_T^{e}\rangle|$ for $B_{(s)}$ decays. More precisely, one has $\langle P_T^{\mu}\rangle/\langle P_T^{e}\rangle\sim10^{2}$ for $D_{(s)}$ decay and $\langle P_T^{\mu}\rangle/\langle P_T^{e}\rangle\sim10^{3}$ for $B_{(s)}$ decay. Again, these ratios correspond to $m_{\mu}/m_{e}$ and $m_{\tau}/m_{e}$, cf. Eq.~\eqref{eqpt}. $\langle P_L^{e}\rangle$ equals 1 for all the cases since in Eq.~\eqref{eqpl}, the term proportional to $\delta_l$ almost vanishes for a numerical calculation. Clearly, in the zero lepton mass limit, $\langle P_T^{l}\rangle=0$ and $\langle P_L^{e}\rangle=1$ for all channels, and $\langle C_F^{e}\rangle=-1.5$ for the $D_{(s)}/B_{(s)}\rightarrow P$ case. In Table $\mathrm{\ref{tabob}}$, the results from other models are still lacking. We note that some of those channels have been explored e.g., in the RQM \cite{Ebert:2006nz,Ebert:2010dv,Faustov:2012nk,Faustov:2014zva,Faustov:2012mt,Faustov:2013ima}, in the QCD sum rule \cite{Azizi:2008tt,Azizi:2008vt},
in the light-cone sum rule \cite{Wang:2015vgv,Wang:2017jow,Lu:2018cfc,Gao:2019lta,Azizi:2010zj}, however, the relevant observables are not given.

Through the longitudinal polarization of the final vector meson, we can obtain the ratios of the partial decay rates $\Gamma_L/\Gamma_T=\langle F_L\rangle/(1-\langle F_L\rangle)$. Experimental measurements give the results regarding these ratios $\Gamma_L/\Gamma_T$ for $D^+\rightarrow \bar{K}^{*0}l^+\nu_l$, and $D_s^+\rightarrow \phi l^+\nu_l$ decays, and we give the comparison in Table~\ref{tabratio}. Our predictions are in good agreement with the available experimental values and other theoretical results from the CCQM~\cite{Ivanov:2019nqd} and RQM~\cite{Faustov:2019mqr}. The PDG average for $\Gamma_L/\Gamma_T$ in $D^+\rightarrow \bar{K}^{*0}l^+\nu_l$ is $1.13\pm 0.08$, which
is fully taken from the measurements for $D^+\rightarrow \bar{K}^{*0}\mu^+\nu_\mu$ with
the exclusion of the data from Ref.~\cite{Anjos:1990pn} for the electron mode.
The PDG average for $\Gamma_L/\Gamma_T$ in $D_s^+\rightarrow \phi l^+\nu_l$ is $0.72\pm0.18$, which is based on the value from Ref.~\cite{Avery:1994hq} as well as
from Refs.~\cite{Frabetti:1994ct} and \cite{Kodama:1993bb}. The last two treat $\Gamma_L/\Gamma_T$ for a lepton mass of zero.

\begin{table}[!htbp]
\centering
\caption{The ratios of the partial decay rates $\Gamma_L/\Gamma_T$ for $D^+\rightarrow \bar{K}^{*0}l^+\nu_l$ and $D_s\rightarrow \phi l^+\nu_l$ decays. The PDG averages are discussed in the text.}
\begin{tabular}{!{\vrule width1pt}c!{\vrule width1pt}p{2cm}<{\centering}p{2cm}<{\centering}p{2cm}<{\centering}c!{\vrule width1pt}}
\Xhline{1pt}
\;&CLFQM&CCQM~\cite{Ivanov:2019nqd}&RQM~\cite{Faustov:2019mqr}&Experimental\\
\Xhline{1pt}
$D^+\rightarrow \bar{K}^{*0}e^+\nu_e$&$1.21$&$1.17$&$1.17$&\space\\
$D^+\rightarrow \bar{K}^{*0}\mu^+\nu_{\mu}$&1.17&1.08&1.17&$1.13\pm0.08$~\cite{Zyla:2020zbs}\\
$D_s\rightarrow \phi e^+\nu_e$&$1.17$&$1.12$&$1.17$&$1.0\pm0.3\pm0.2$~\cite{Avery:1994hq}\\
$D_s\rightarrow \phi \mu^+\nu_{\mu}$&$1.12$&$1$&$1.17$&\space\\
\Xhline{1pt}
\end{tabular}
\label{tabratio}
\end{table}

Considering the $l^-\bar{\nu}_l$ mode, we recalculate the relevant physical observables for the $\bar{B}^0\rightarrow D^+l^-\bar{\nu}_l$ and $\bar{B}^0\rightarrow D^{*+}l^-\bar{\nu}_l$ decay channels and compare our results with another prediction from the CCQM~\cite{Ivanov:2015tru} in Table \ref{tabminus}. Again there is a good agreement.

\begin{table}[!htbp]
\centering
\caption{Forward-backward asymmetry, lepton polarization, and convexity parameters for semileptonic decays of $\bar{B}^0\rightarrow D^+l^-\bar{\nu}_l$ and $\bar{B}^0\rightarrow D^{*+}l^-\bar{\nu}_l$, compared with the prediction from the CCQM~\cite{Ivanov:2015tru},
where $\langle P_T^e\rangle$ is shown as 0 since it is on the order of $10^{-3}$. }

\begin{tabular}{!{\vrule width1pt}c!{\vrule width1pt}ccccccccc!{\vrule width1pt}}
\Xhline{1pt}
\;&\;&$\langle\mathcal{A}_{FB}^e\rangle$&$\langle\mathcal{A}_{FB}^\tau\rangle$&$\langle P_L^e\rangle$&$\langle P_L^\tau\rangle$&$\langle P_T^e\rangle$&$\langle P_T^\tau\rangle$&$\langle C_F^e\rangle$&$\langle C_F^\tau\rangle$\\
\Xhline{1pt}
\multirow{2}*{$\bar{B}^0\rightarrow D^+l^-\bar{\nu}_l$}&
CLFQM&$-1.04\times10^{-6}$&$-0.36$&$-1$&$0.32$&$1.06\times10^{-3}$&$0.84$&$-1.5$&$-0.27$\\
&CCQM&$-1.17\times10^{-6}$&$-0.36$&$-1$&$0.33$&$\;$&$0.84$&$-1.5$&$-0.26$\\
\hline
\multirow{2}*{$\bar{B}^0\rightarrow D^{*+}l^-\bar{\nu}_l$}&
CLFQM&$0.22$&$0.054$&$-1$&$-0.51$&$0.46\times10^{-3}$&$0.47$&$-0.42$&$-0.056$\\
&CCQM&$0.19$&$0.027$&$-1$&$-0.50$&$\;$&$0.46$&$-0.47$&$-0.062$\\
\Xhline{1pt}
\end{tabular}
\label{tabminus}
\end{table}

\begin{table}[!htbp]
\centering
\caption{The average values for trigonometric moments (see Eq.(\ref{eqw})), compared with the prediction from the CCQM \cite{Ivanov:2019nqd}.}
\begin{tabular}{!{\vrule width 1pt}ccp{1.4cm}<{\centering}p{1.4cm}<{\centering}p{1.4cm}<{\centering}p{1.4cm}<{\centering}p{1.4cm}<{\centering}p{1.4cm}<{\centering}!{\vrule width 1pt}}
\Xhline{1pt}
\;&\;&$\langle W_{T}^e\rangle$&$\langle W_{T}^{\mu}\rangle$&$\langle W_I^e\rangle$&$\langle W_I^{\mu}\rangle$&$\langle W_A^e\rangle$&$\langle W_A^{\mu}\rangle$\\
\Xhline{1pt}
$D^+\rightarrow\rho l^+\nu_l$&CLFQM&$-0.078$&$-0.077$&$0.052$&$0.049$&$-0.078$&$-0.072$\\
\;&CCQM&$-0.091$&$-0.089$&$0.054$&$0.051$&$-0.067$&$-0.061$\\

$D^+\rightarrow\omega l^+\nu_l$&CLFQM&$-0.081$&$-0.080$&$0.052$&$0.049$&$-0.077$&$-0.071$\\
\;&CCQM&$-0.093$&$-0.091$&$0.054$&$0.051$&$-0.066$&$-0.060$\\

$D^+\rightarrow\bar{K^*}l^+\nu_l$&CLFQM&$-0.091$&$-0.089$&$0.054$&$0.051$&$-0.062$&$-0.055$\\
\;&CCQM&$-0.097$&$-0.094$&0.055&0.051&$-0.056$&$-0.049$\\
\hline
$D_s^+\rightarrow\phi l^+\nu_l$&CLFQM&$-0.095$&$-0.093$&$0.054$&$0.051$&$-0.057$&$-0.050$\\
\;&CCQM&$-0.101$&$-0.098$&$0.055$&$0.052$&$-0.055$&$-0.048$\\

$D_s^+\rightarrow K^* l^+\nu_l$&CLFQM&$-0.085$&$-0.083$&$0.053$&$0.050$&$-0.072$&$-0.066$\\
\;&CCQM&$-0.094$&$-0.092$&$0.054$&$0.051$&$-0.068$&$-0.062$\\
\Xhline{1pt}
\;&\;&$\langle W_{T}^e\rangle$&$\langle W_{T}^{\tau}\rangle$&$\langle W_I^e\rangle$&$\langle W_I^{\tau}\rangle$&$\langle W_A^e\rangle$&$\langle W_A^{\tau}\rangle$\\
\Xhline{1pt}
$B^+\rightarrow\rho l^+\nu_l$&CLFQM&$-0.050$&$-0.042$&$0.048$&$0.031$&$-0.109$&$-0.056$\\
$B^+\rightarrow\omega l^+\nu_l$&CLFQM&$-0.056$&$-0.048$&$0.049$&$0.032$&$-0.103$&$-0.058$\\
$B^+\rightarrow\bar{D^*}l^+\nu_l$&CLFQM&$-0.091$&$-0.056$&$0.054$&$0.025$&$-0.069$&$-0.005$\\
\hline
$B_s\rightarrow K^{*-} l^+\nu_l$&CLFQM&$-0.063$&$-0.053$&$0.051$&$0.034$&$-0.085$&$-0.042$\\
$B_s\rightarrow D_s^{*-} l^+\nu_l$&CLFQM&$-0.092$&$-0.056$&$0.054$&$0.025$&$-0.067$&$-0.043$\\
\Xhline{1pt}
\end{tabular}
\label{tabw}
\end{table}

In Table $\mathrm{\ref{tabw}}$, we present our predictions for the trigonometric moments of $D_{(s)}$ and $B_{(s)}$ decay. Except for the electron mode, for $D_{(s)}$ decay, we have also calculated the muon mode, while for $B_{(s)}$, the tau mode is also illustrated. For the transitions of meson $D_{(s)}$, we give the comparison with the predictions of the CCQM \cite{Soni:2018adu}, and we find that our predictions are in good agreement with them.

As a drawback of the model, it is hard to quantify the theoretical uncertainty.
Comparing the central values between Refs.~\cite{Verma:2011yw} and \cite{Cheng:2003sm}, one finds that a general 7\%-10\% uncertainly attached to form factor is reasonable. Reference \cite{Verma:2011yw} is just an update of \cite{Cheng:2003sm} with utilizing the latest experimental information, wherever available, or the lattice results to fix the parameter $\beta$ in the wave function of meson involved. Considering also the uncertainty  from CKM matrix elements
(the uncertainty for $|V_{ub}|$ is larger due to the difference between the extractions from the exclusive and inclusive mode), a 10\%-15\% uncertainty will be assigned to the observable.

\begin{table}[!htbp]
\centering
\caption{Our predictions for $F_L^{\tau}(D_{(s)}^{*})$ and $P_L^{\tau}(D_{(s)}^{(*)})$, compared with other models as well as experimental values. In parenthesis, we also include the value of $F_{L}^e(D^*)$ for $\bar{B}\rightarrow D^*\tau^{-}\bar{\nu}_{\tau}$.}
\begin{tabular}{!{\vrule width 1pt}c|c|cc|cc!{\vrule width 1pt}}
\Xhline{1pt}
Observables&Approach&$\bar{B}\rightarrow D\tau^{-}\bar{\nu}_{\tau}$&$\bar{B}\rightarrow D^*\tau^{-}\bar{\nu}_{\tau}\,(e^{-}\bar{\nu}_{e})$&$B_s\rightarrow D_s\tau^{-}\bar{\nu}_{\tau}$&$B_s\rightarrow D_s^*\tau^{-}\bar{\nu}_{\tau}$\\
\Xhline{1pt}
\multirow{6}*{$F_L^{\tau}(D_{(s)}^{*})$}&CLFQM&-&0.451 (0.521)&-&0.453\\
&SM1 \cite{Alok:2016qyh}&-&$0.46\pm0.04$&-&-\\
&SM2 \cite{Das:2019cpt}&-&0.455&-&0.433\\
&PQCD \cite{Hu:2019bdf}&-&0.43&-&0.43\\
&Belle \cite{Abdesselam:2019wbt} &-&$0.60\pm0.08\pm0.04$ ($0.56\pm0.02$) &-&-\\
\Xhline{1pt}
\multirow{6}*{$P_L^{\tau}(D_{(s)}^{(*)})$}&CLFQM&$0.32$&$-0.51$&$0.33$&$-0.51$\\
&SM1 &$0.325\pm0.09$ \cite{Tanaka:2010se}&$-0.497\pm0.013$ \cite{Tanaka:2012nw}&-&-\\
&SM2 \cite{Das:2019cpt}&0.352&$-0.501$&-&$-0.520$\\
&PQCD \cite{Hu:2019bdf}&0.30&$-0.53$&0.30&$-0.53$\\
&Belle \cite{Hirose:2016wfn}&-&$-0.38\pm0.51^{+0.21}_{-0.16}$&-&-\\
\Xhline{1pt}
\end{tabular}
\label{tabc}
\end{table}

The Belle collaboration has measured the longitudinal polarization of the final vector meson as $F_L^{\tau}(D^*)=0.60\pm0.08\pm0.04$ \cite{Abdesselam:2019wbt} and the longitudinal polarization of the $\tau$ lepton\footnote{Notice that the Belle collaboration measures the longitudinal polarization of lepton with $\tau$ decays $\tau\rightarrow \pi^-\nu_{\tau}$ and $\tau\rightarrow \rho^-\nu_{\tau}$ in the rest frame of $\tau$. In fact, it could also be measured in laboratory frame of $\tau$ or in the rest frame of virtual W \cite{Hagiwara:1989fn}. Nevertheless, in the rest frame of $\tau$, the formula becomes much simpler.} as $P_L^{\tau}(D^*)=-0.38\pm0.51^{+0.21}_{-0.16}$ \cite{Hirose:2016wfn} for the decay $\bar{B}\rightarrow D^*\tau^-\nu_{\tau}$. Our results from the CLFQM, as well as other predictions, are compared to those from the experiment in Table \ref{tabc}. The results agree well within uncertainties; however, the uncertainty for $P_L^{\tau}$ of the Belle measurement is very large. Further precise measurement is important and desirable. As discussed in many works \cite{Hirose:2016wfn,Blanke:2018yud,Shi:2019gxi,Ivanov:2020iad,Alok:2018uft,Alok:2019uqc}, the polarization observables $F_L^{\tau}(D^*)$, $P_L^{\tau}(D^*)$, and the yet unmeasured $P_L^{\tau}(D)$ could potentially discriminate the effects of new operator structure beyond the SM. Their correlation with other NP models---the two Higgs doublet model and a leptoquark model---are discussed in Refs.~\cite{Tanaka:2010se,Tanaka:2012nw}. We also notice that the values of $P^\tau$ and $\mathcal{A}_{FB}$ in Table ${\mathrm{\uppercase\expandafter{\romannumeral6}}}$ in Ref.~\cite{Hu:2019bdf} should correspond to the $\tau^-\bar{\nu}_{\tau}$ mode, not the $\tau^+\nu_{\tau}$ mode. As mentioned, the two modes will lead to different results with not only an overall sign difference, cf. Eq.~\eqref{eqp} and Eqs.~\eqref{eqafb},~\eqref{eqpl}, and~\eqref{eqpt}.

As a matter of fact, there are very different landscapes for polarizations of $\tau$ and lighter ones $e$ and $\mu$. The polarisation of $\tau$ can be accessed by analysing its decay product, as has been done by Belle collaboration, but it is not possible for $e$  and $\mu$. In a complete angular analysis, cf. Eq.~(40), the helicity structure functions $\mathcal{H}_{U,L,S,P,SL}$ could be extracted, and from them the polarisation observables are defined.

\begin{figure}[ht]
\begin{minipage}{1.0\linewidth}
\centerline{\includegraphics[width=0.45\textwidth]{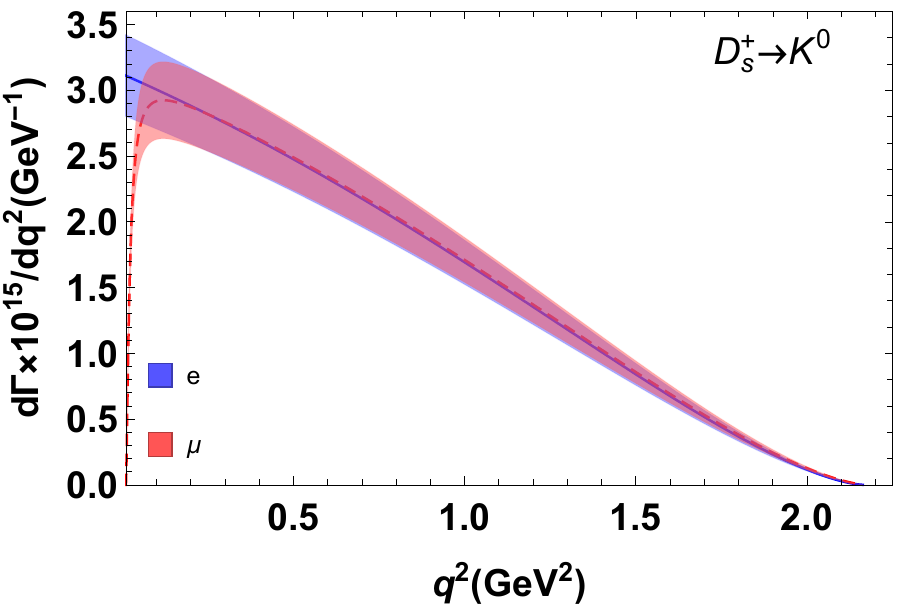}\;\;\;\;\;\;\includegraphics[width=0.45\textwidth]{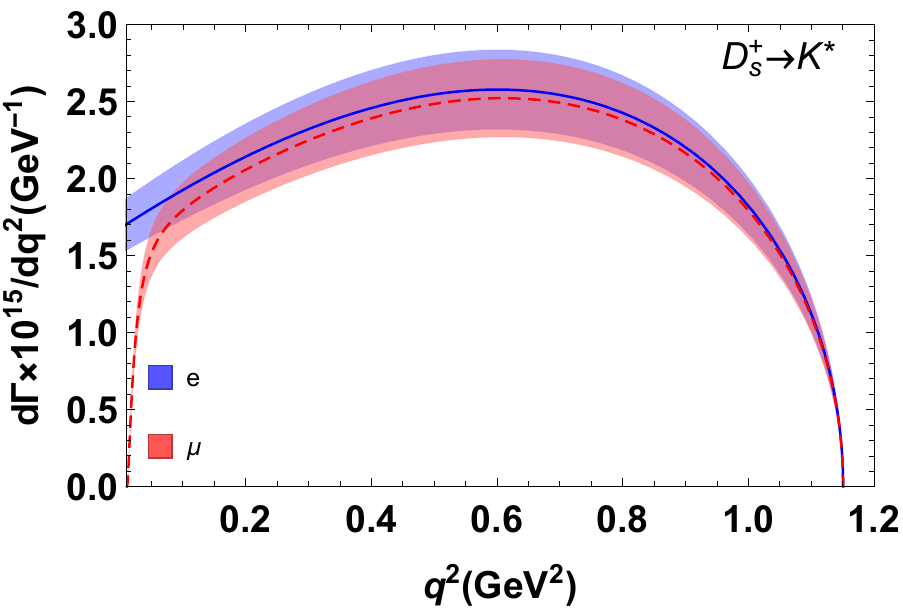}}
\centerline{\includegraphics[width=0.45\textwidth]{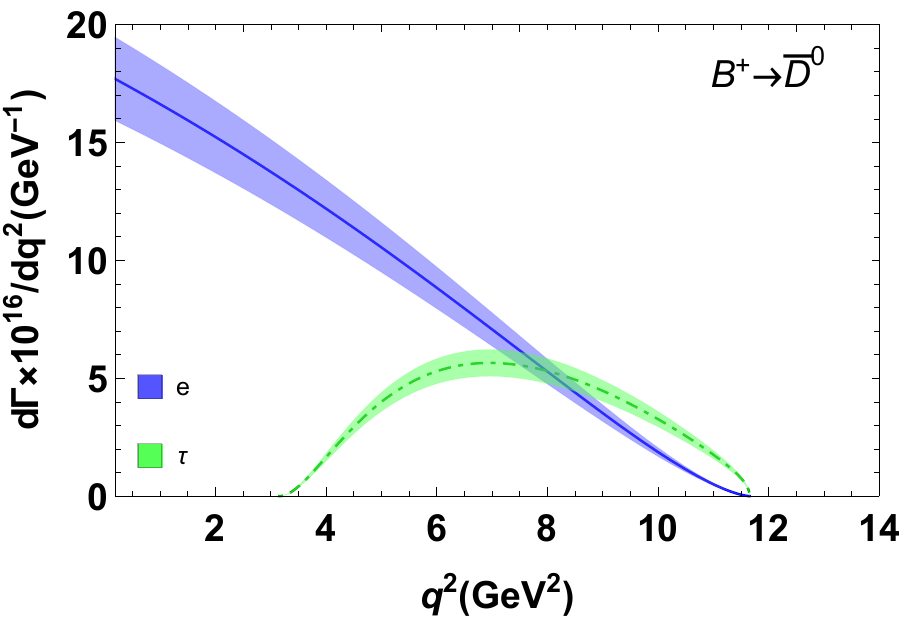}\;\;\;\;\;\;\includegraphics[width=0.45\textwidth]{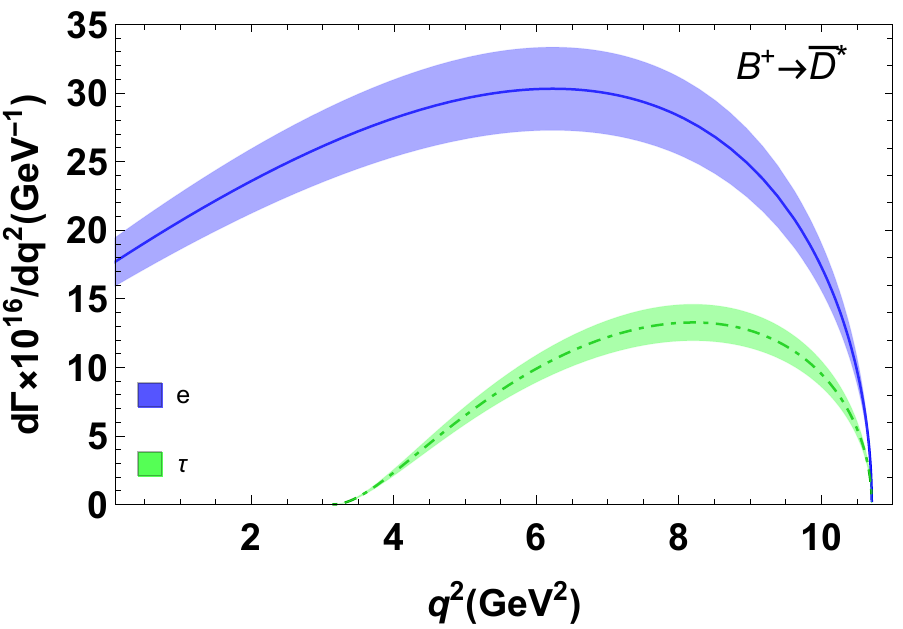}}
\centerline{\includegraphics[width=0.45\textwidth]{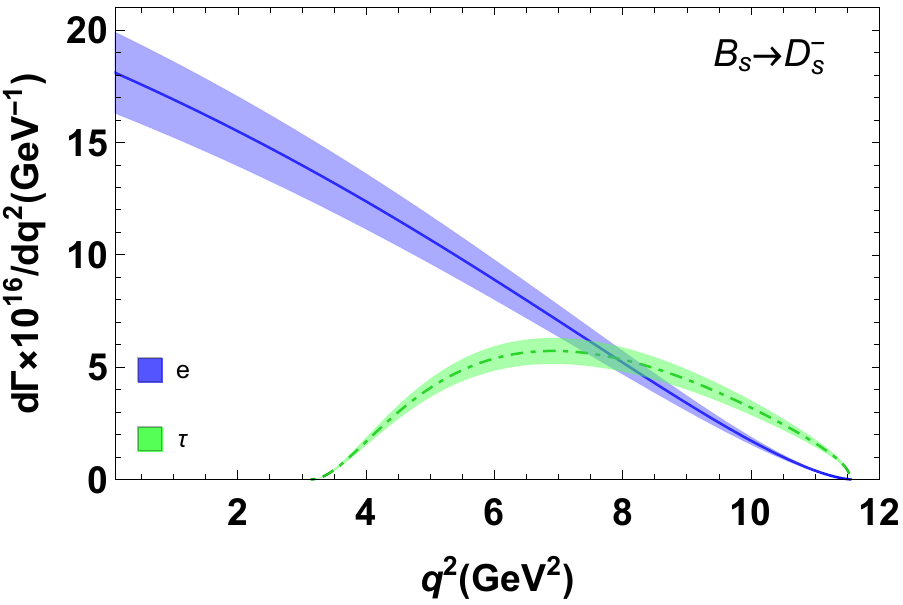}\;\;\;\;\;\;\includegraphics[width=0.45\textwidth]{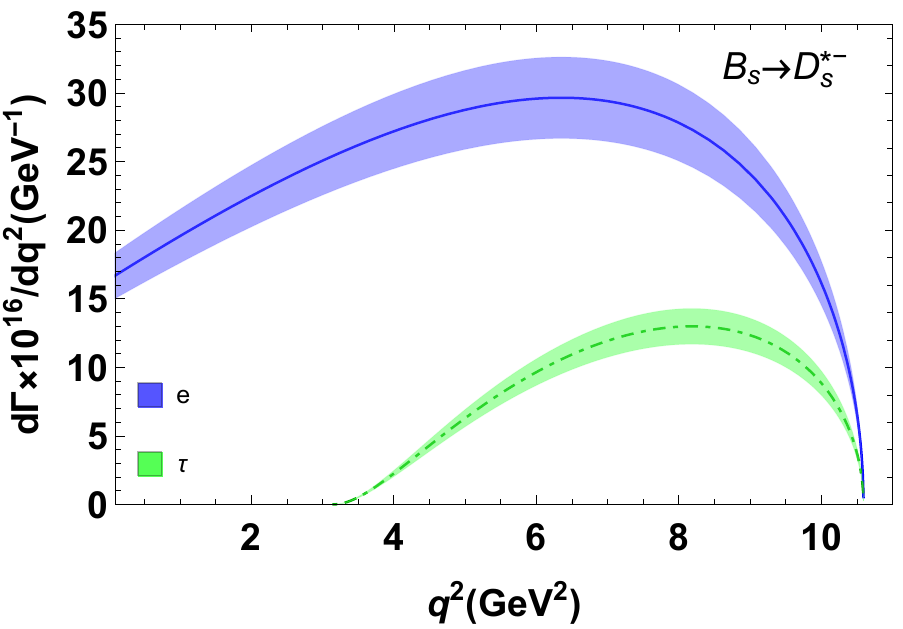}}
\end{minipage}
\caption{The differential decay rates of the decays $D_s^+\rightarrow K^{(*)}l^+\nu_l$, $B^+\rightarrow \bar{D}^{(*)}l^+\nu_l$ and $B_s\rightarrow D_s^{(*)-}l^+\nu_l$. The lepton mode is indicated by the corresponding legend. The solid and dashed lines denote the central values and the band demonstrates
our estimated uncertainty.}
\label{figbr}
\end{figure}

We also show the differential decay distribution $d\Gamma/dq^2$ within the full range of the momentum transfer squared in Fig.~\ref{figbr} for some selected channels. These are direct observables in the experiment and will be tested in the future.

\begin{figure}[ht]
\begin{minipage}{1.0\linewidth}
\centerline{\includegraphics[width=0.5\textwidth]{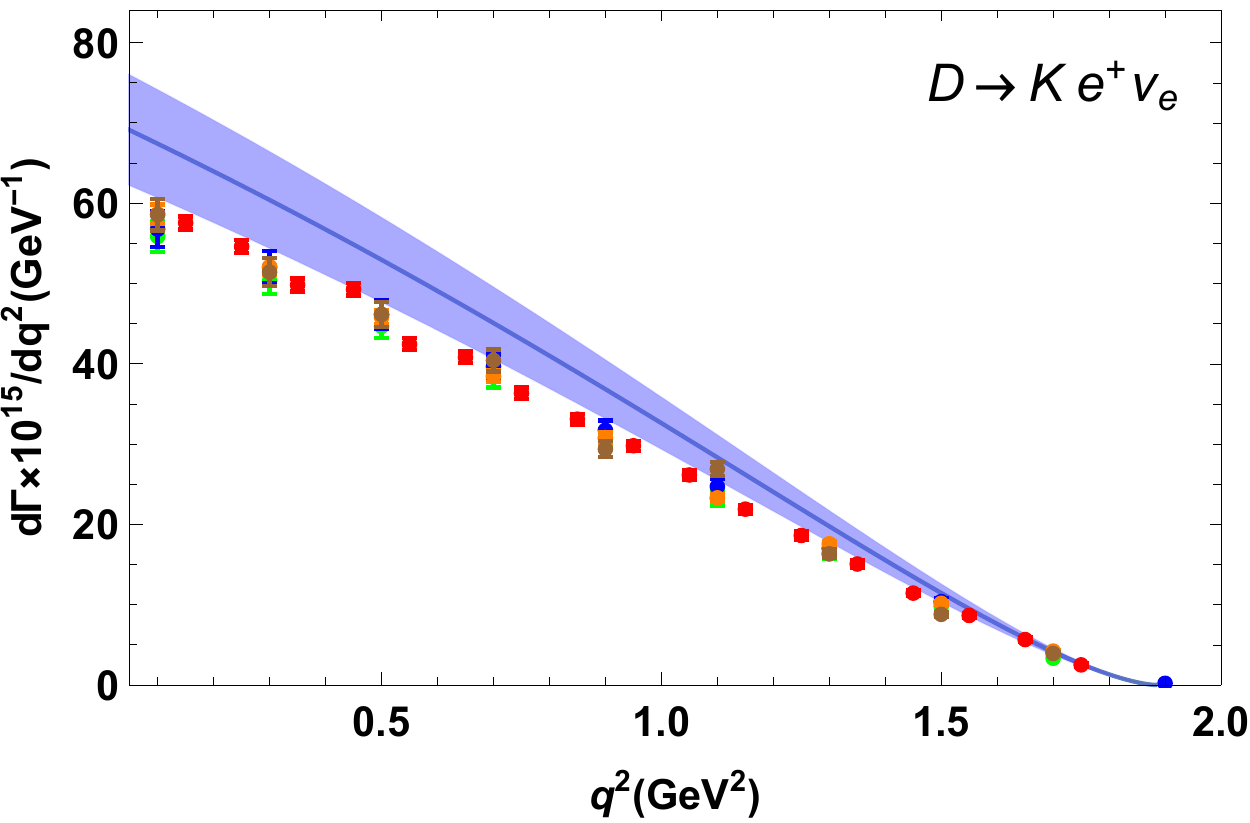}\;\;\;\;\;\;\includegraphics[width=0.5\textwidth]{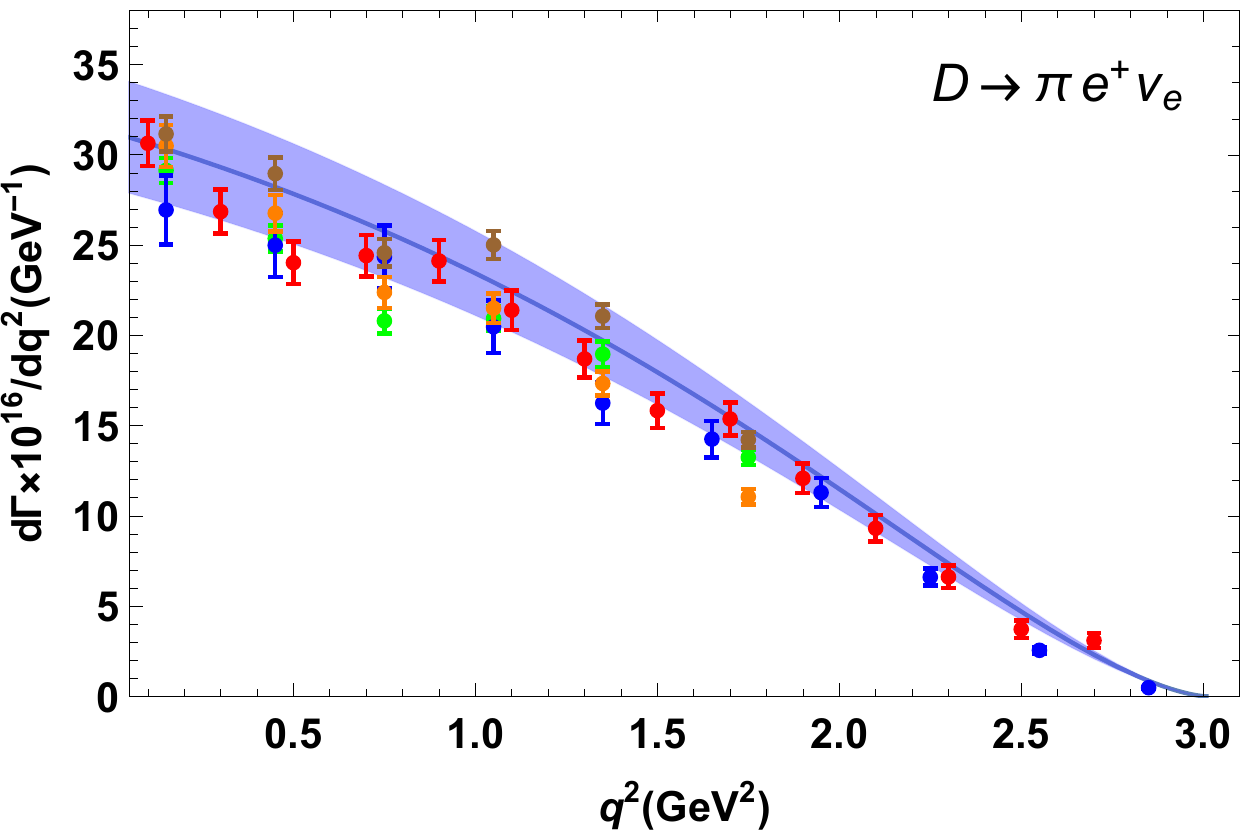}}
\end{minipage}
\caption{The differential decay rate for the decays $D\rightarrow Ke^{+}\nu_e$ and $D\rightarrow \pi e^+ \nu_e$. The solid line indicates our central values and the band indicates the estimated uncertainty. We have used the experimental data from BES ${\mathrm{\uppercase\expandafter{\romannumeral3}}}$ for neutral $D^0$ \cite{Ablikim:2015ixa} (red dots with error bars) and charged $D^+$ \cite{Ablikim:2017lks} (green dots with error bars), BaBar \cite{Lees:2014ihu,Aubert:2007wg} (blue dots and error bars) and CLEO \cite{Besson:2009uv} for neutral $D^0$ (orange dots and error bars) and charged $D^+$ (brown dots with error bars).}
\label{figc}
\end{figure}

Moreover, we compare our results with the experimental results of the differential decay rate for $D\rightarrow Ke^+\nu_e$ and $D\rightarrow \pi e^+\nu_e$ in Fig.~\ref{figc}. The experimental data are from the BES${\mathrm{\uppercase\expandafter{\romannumeral3}}}$ \cite{Ablikim:2015ixa,Ablikim:2017lks}, BABAR \cite{Lees:2014ihu,Aubert:2007wg} and CLEO \cite{Besson:2009uv} collaboration. The blue band is obtained by assigning the central values a $10\%$ uncertainty, demonstrating to some extent our theoretical uncertainty. Our results for $D\rightarrow \pi e^+\nu_e$ agree very well with the experimental findings, while for the $D\rightarrow Ke^+\nu_e$ case, they only agree within their uncertainties. As already noted in Ref.~\cite{Faustov:2019mqr}, our result for $D\rightarrow Ke^+\nu_e$ is larger than theirs and the experiment.

\begin{figure}[ht]
\begin{minipage}{1.0\linewidth}
\centerline{\includegraphics[width=1.2\textwidth]{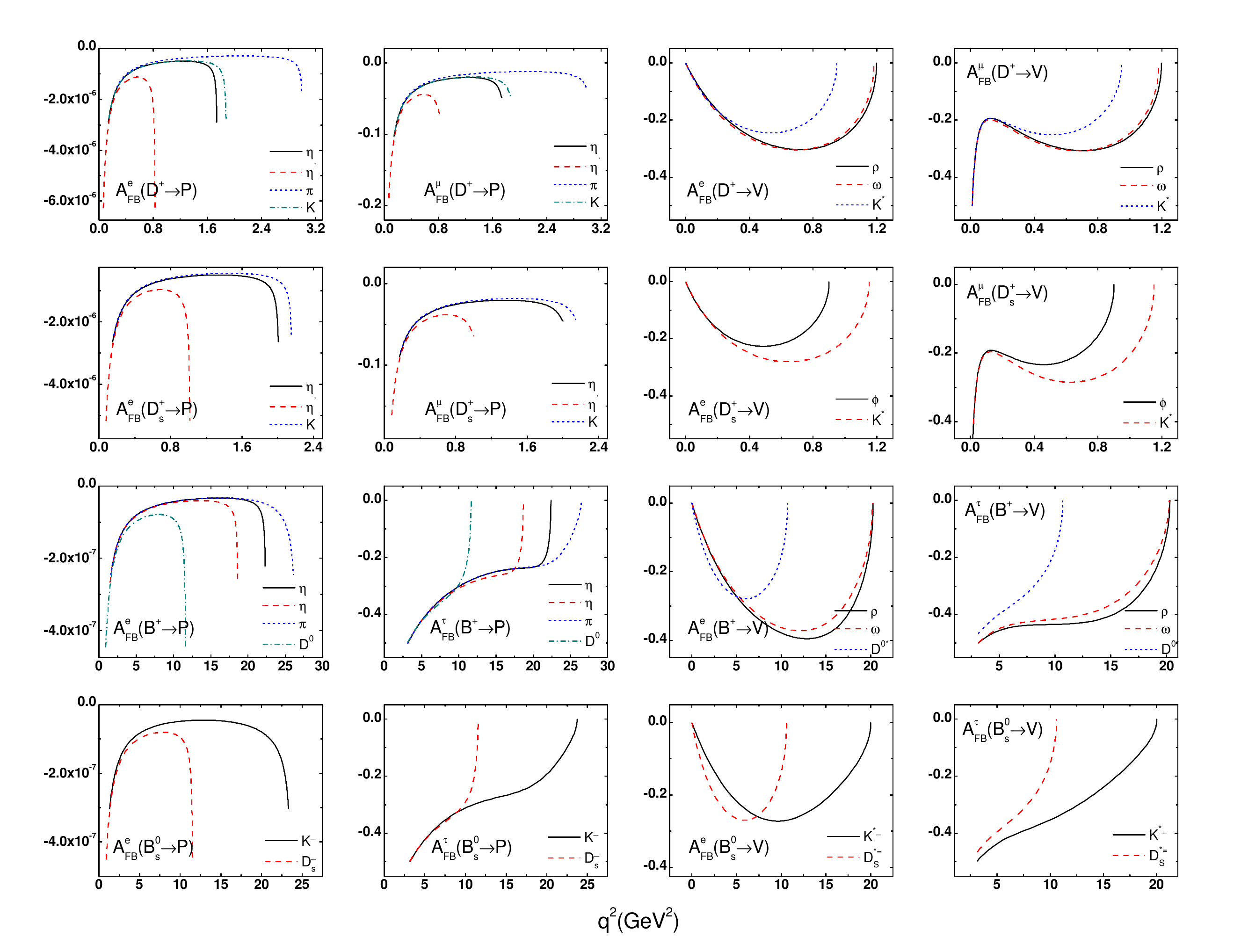}}
\centerline{}
\end{minipage}
\caption{The forward-backward asymmetries of the decays $D_{(s)}\rightarrow P(V)l^+\nu_l$ and $B_{(s)}\rightarrow P(V)l^+\nu_l$.}
\label{figa}
\end{figure}

\begin{figure}[ht]
\begin{minipage}{1.0\linewidth}
\centerline{\includegraphics[width=1.2\textwidth]{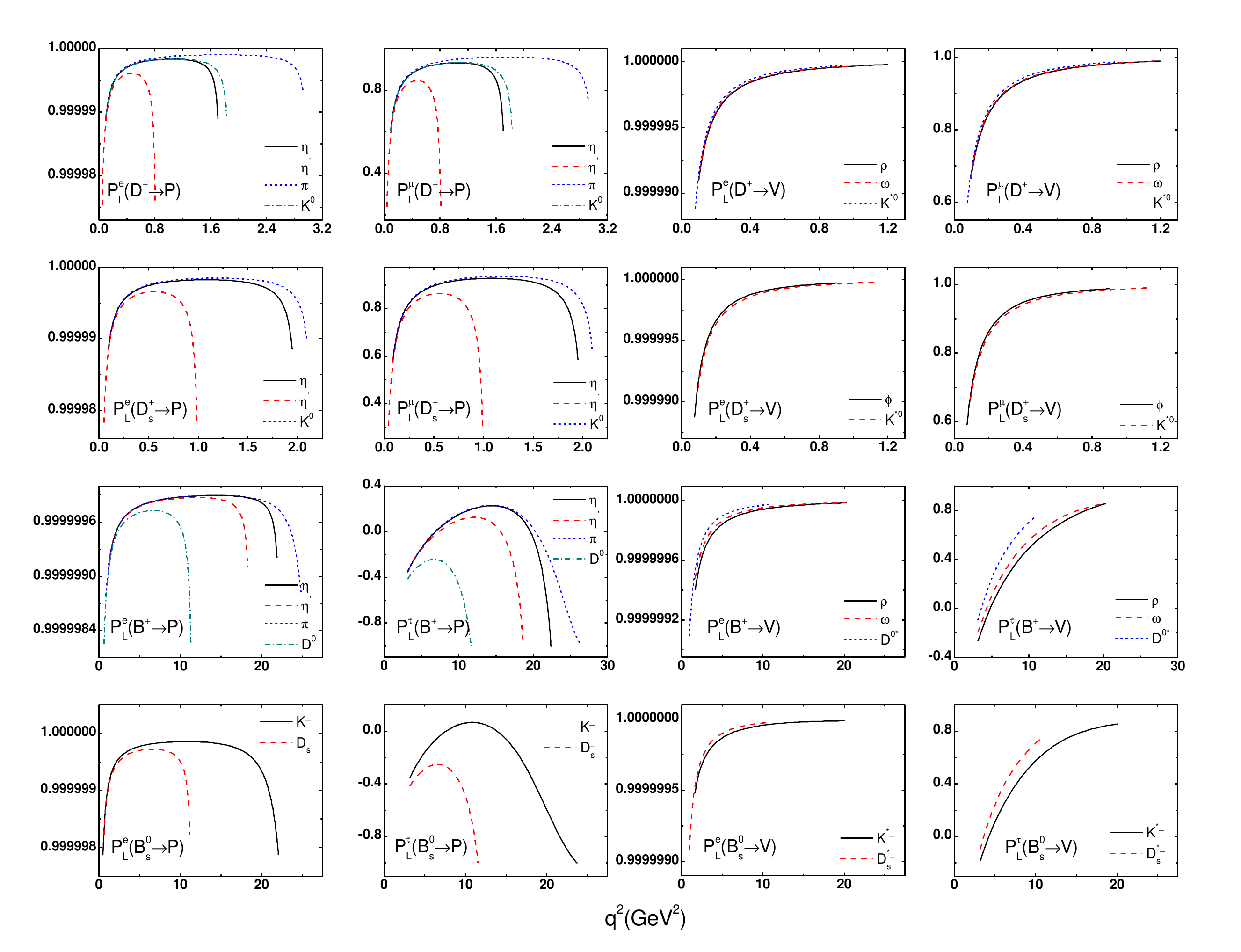}}
\centerline{}
\end{minipage}
\caption{The longitudinal polarization of a charged lepton of the decays $D_{(s)}\rightarrow P(V)l^+\nu_l$ and $B_{(s)}\rightarrow P(V)l^+\nu_l$.}
\label{figpl}
\end{figure}

\begin{figure}[ht]
\begin{minipage}{1.0\linewidth}
\centerline{\includegraphics[width=1.2\textwidth]{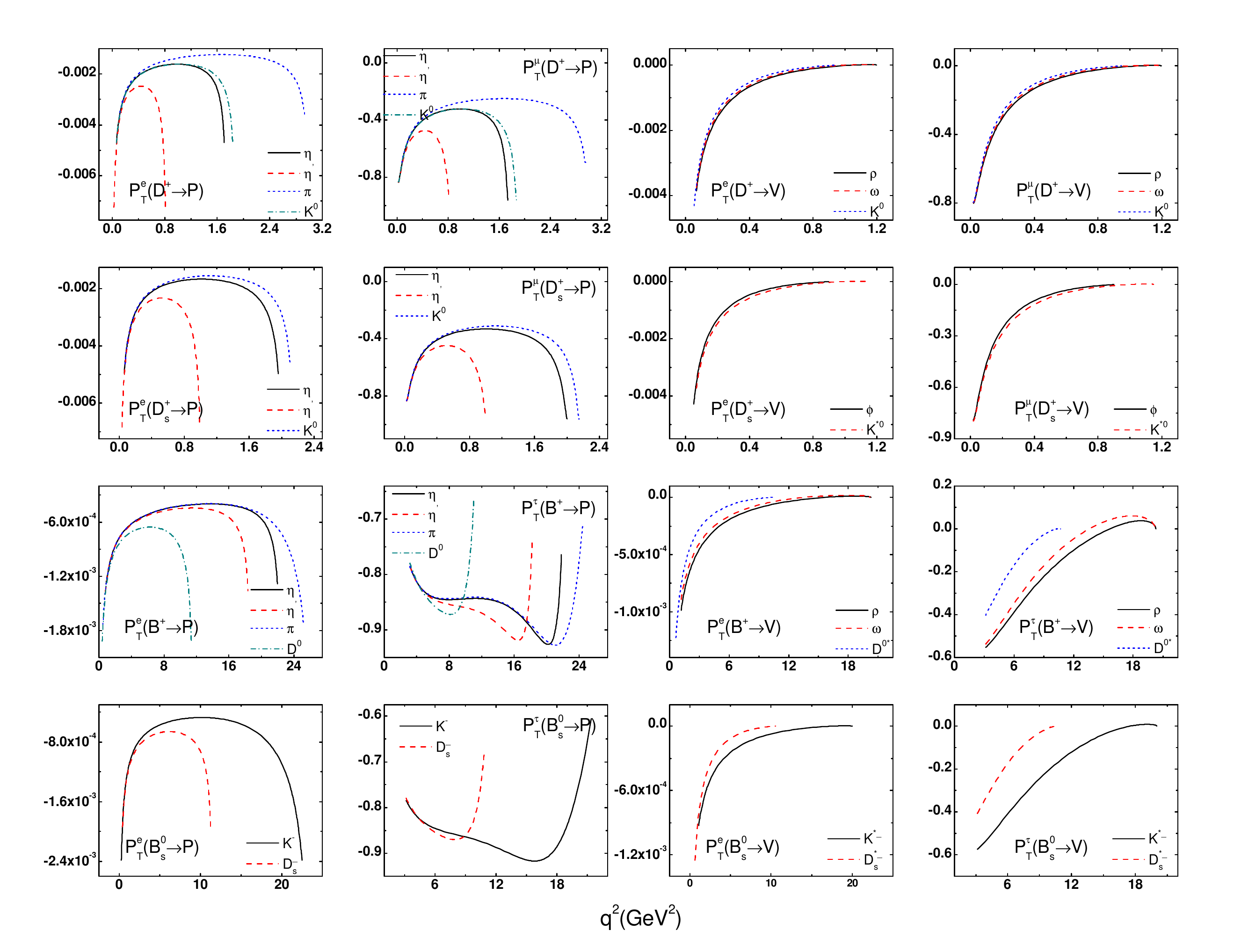}}
\centerline{}
\end{minipage}
\caption{The transverse polarization of a charged lepton of the decays $D_{(s)}\rightarrow P(V)l^+\nu_l$ and $B_{(s)}\rightarrow P(V)l^+\nu_l$.}
\label{figpt}
\end{figure}

In Figs.~\ref{figa}-\ref{figpt}, we represent the $q^2$ dependence of the forward-backward asymmetry and the leptonic longitudinal and transverse polarization, respectively. For the decays $D_{(s)}\rightarrow Vl^+\nu_l$ and $B_{(s)}\rightarrow Vl^+\nu_l$, we can find that the values of $\mathcal{A}_{FB}$ and $P_T$ coincide with 0 at the zero recoil point ($q^2=q_\text{max}^2$) since $\mathcal{H}_P=0$, $\mathcal{H}_{SL}=0$, cf. Eq.~\eqref{eqafb} and Eq.~\eqref{eqpt}.
Owing to the same reason, the leptonic longitudinal polarization Eq.~\eqref{eqpl} for the mode of the vector meson in final states is reduced to $(1-\delta_l)/(1+\delta_l)$ at the zero recoil point; then, the values of $P_L^{e(\mu)}$ approach 1, as shown in Fig.~\ref{figpl}.
In all these figures, one finds that the longitudinal and transverse polarization values for $\mu$ and $\tau$ are larger than the electron, which illustrates the lepton mass effect as expected.

\clearpage
\section{SUMMARY}
\label{Sec:summary}
Based on the form factors computed from the covariant light front quark model \cite{Cheng:2003sm,Verma:2011yw}, we perform a comprehensive analysis of the semileptonic decays $D_{(s)},B_{(s)}\rightarrow P(V)l\nu_l$ in the helicity formalism, where all the observables are expressed by form factors via helicity amplitude. We provide the detailed derivation for the differential decay rate $(d\Gamma/dq^2)$, forward-backward asymmetry $(\mathcal{A}_{FB}^l)$, longitudinal $(P_L^l)$ and transverse polarization $(P_T^l)$ of a lepton, longitudinal polarization of a vector meson $(F_L^l(V))$, leptonic convexity parameter $(C_F^l)$, and trigonometric moments $(W_L^l, W_I^l,W_A^l)$. The numerical results are shown in Sec.~\ref{Sec:results}.

The values of the branching fractions are in good agreement with the experiment for $D_{(s)}$ and $B$ decay; however, the experimental information is lacking for $B_s$ decay. For the polarization observables, the only measurements are $F_L^{\tau}(D^*)$ and $P_L^{\tau}(D^*)$ for the decay $\bar{B} \rightarrow D^*\tau^-\nu_{\tau}$ from the Belle collaboration. The current Standard Model prediction agrees with the experimental value for the polarization of $D^*$ ($F_L^{\tau}(D^*)$), within 1.6 standard deviations of the mean, while the experimental values for the polarization of $\tau$ ($P_L^{\tau}(D^*)$) suffer from large uncertainty. As recognized by our community, these polarization observables are crucial inputs for testing and investigating New Physics. Thus, further examination from both the theoretical and experimental sides is of great importance.

These observables could be measured in BES${\mathrm{\uppercase\expandafter{\romannumeral3}}}$, Belle, and LHCb. From a practical perspective, the forward-backward asymmetry could be measured as a first step. In the near future, Belle${\mathrm{\uppercase\expandafter{\romannumeral2}}}$ will accumulate data samples 50 times as large as those for Belle, which will provide a major opportunity for such further measurements and the determination of variously unmeasured observables discussed in the current work. Concerning the charm part, the super $\tau$-charm factory is planned to be built in China.

\acknowledgments
 The author XWK acknowledges very useful discussions with Profs. M.~A.~Ivanov and V.~Galkin.
 The author XWK is supported by the National Natural Science Foundation of China (NSFC) under Project No. 11805012 and the Fundamental Research Funds for the Central Universities (BNU). XHG acknowledges NSFC under Grant No. 11775024. LYD acknowledges NSFC with Grant No. 11805059 and Joint Large Scale Scientific Facility Funds of the NSFC and Chinese Academy of Sciences (CAS) under Contract No. U1932110, and Fundamental Research Funds for the Central Universities (HNU). CW acknowledges NSFC No. 11805153. TL acknowledges NSFC with Project Nos. 11805037 and U1832121 and Shanghai Pujiang Program under Grant No. 18PJ1401000 and Open Research Program of Large Research Infrastructures (2017), CAS.
\section*{References}
\bibliographystyle{apsrev}
\bibliography{ref4}

\end{document}